\documentclass[preprint]{imsart}

\RequirePackage[OT1]{fontenc}
\RequirePackage{amsthm,amsmath}
\RequirePackage[numbers]{natbib}
\RequirePackage[colorlinks,citecolor=blue,urlcolor=blue]{hyperref}

\startlocaldefs
\numberwithin{equation}{section}
\theoremstyle{plain}

\endlocaldefs

\usepackage{psfrag,
graphicx, 
epsf,yhmath,
euscript,amsfonts,amsmath,latexsym,amssymb,mathrsfs
}
\usepackage[active]{srcltx}
\usepackage{mathabx}

\newtheorem{lemma}{Lemma}
\newtheorem{theorem}{Theorem}

\newtheorem{example}{Example}

\newtheorem{remark}{Remark}
\newtheorem{definition}{Definition}

\newcommand{\cF}{{\cal F}}

\newcommand{\cL}{{\cal L}}
\newcommand{\cM}{{\cal M}}

\newcommand{\cR}{{\cal R}}

\newcommand{\bC}{\mathbb C}

\newcommand{\bE}{\mathbb E}

\newcommand{\bL}{{\mathbb L}}

\newcommand{\bN}{{\mathbb N}}

\newcommand{\bR}{{\mathbb R}}

\newcommand{\sF}{{\mathscr F}}

\newcommand{\sH}{{\mathscr H}}

\newcommand{\rd}{\mathrm{d}}

\renewcommand{\kappa}{\varkappa}

\newcommand{\epr}{\hfill\hbox{\hskip 4pt
                \vrule width 5pt height 6pt depth 1.5pt}\vspace{0.5cm}\par}
\begin{document}

\begin{frontmatter}
\title{Nonparametric density estimation from observations with multiplicative measurement errors\thanksref{T1}}
\thankstext{T1}{The research is supported by the Russian Academic Excellence Project ``5-100''  and by the Israel Science Foundation (ISF)  research grant 361/15.}
\runtitle{Density estimation under multiplicative errors}

\begin{aug}
\author{\fnms{Denis} \snm{Belomestny}\thanksref{t1, t3} \ead[label=e1]{denis.belomestny@uni-due.de}} 
\and
\author{\fnms{Alexander} \snm{Goldenshluger}\thanksref{t2}\ead[label=e2]{goldensh@stat.haifa.ac.il}}

\thankstext{t1}{Faculty of Mathematics,
Duisburg-Essen University,
D-45127 Essen,
Germany.   \printead{e1}
}
\thankstext{t2}{Department of Statistics, University of Haifa, Haifa 31905, Israel. \printead{e2}} 
\thankstext{t3}{National Research University 
Higher School of Economics, Moscow.}
\runauthor{D. Belomestny and A. Goldenshluger}

\affiliation{Duisburg-Essen University\thanksmark{m1} and University of Haifa\thanksmark{m2}}


\end{aug}

%

\begin{abstract} 
In this paper we study the problem of 
pointwise density estimation from observations with multiplicative 
measurement errors. We elucidate  the main feature of this problem: 
the influence of the estimation point  on the estimation accuracy.
In particular, we show that,  depending on whether this point is separated away from zero or not,
there are two different regimes in terms of the  rates of convergence of the minimax risk. 
In both regimes we develop kernel--type density estimators and prove upper bounds on 
their maximal risk over suitable  nonparametric classes of densities.  We show that the proposed
estimators are rate--optimal by establishing matching lower bounds on the minimax risk. 
Finally we test our estimation procedures on simulated data.
\end{abstract}
 

\end{frontmatter}
\section{Introduction}
\paragraph{Problem formulation and background}
In this paper we study the problem of nonparametric density 
estimation from observations with multiplicative measurements errors. 
In particular, 
assume that  we observe a sample $Y_1,\ldots,Y_n$  generated by the model 
\begin{equation}\label{eq:observations}
 Y_i=X_i\eta_i,\;\;\;i=1,\ldots, n,
\end{equation}
where $X_1,\ldots, X_n$ are independent identically distributed (i.i.d) random variables with density $f_X$,
and 
$\eta_1,\ldots,\eta_n$ are i.i.d. random variables,  independent of $X_1,\ldots, X_n$, 
with known density $g$.
Our goal is to estimate the value of $f_X$ at a single given point $x_0$ from observations $Y_1,\ldots, Y_n$.
If $f_Y$ stands for the density of $Y=X\eta,$ then 
\begin{eqnarray}
 f_Y(y)=[f_X\star g](y)&:=& \int_{-\infty}^\infty \frac{1}{x} f_X(y/x) g(x) \rd x 
 \nonumber
 \\
 &=& \int_{-\infty}^\infty \frac{1}{x} g(y/x) f_X(x)\rd x ,\;\;\;y\in \bR.
\label{eq:f_Y}
\end{eqnarray}
Thus $f_Y$ is a scale mixture of $g$, and estimation of $f_X$ from observations 
$Y_1,\ldots,Y_n$ can be viewed as the problem of  demixing of a scale mixture.  
\par 
The outlined estimation  problem appears in the literature in various  contexts. 
First, the model (\ref{eq:observations}) with normal errors $\eta_i$
and positive random variables $X_i$ represents a 
\textit{stochastic volatility model} without drift. In this context estimation 
of the volatility density $f_X$ from observations $Y_1,\ldots,Y_n$ was studied 
by Van Es et al.~\cite{van2003nonparametric}, Van Es \&~Speij~\cite{van2011estimation} and 
Belomestny \& Shoenmakers~\cite{belomestny2015statistical}.   
\par 
Second, 
if $(\eta_i)$ are uniformly distributed on $[0,1]$ then 
the corresponding model \eqref{eq:observations} 
is referred to as  
the \textit{multiplicative censoring model}.
In this setting 
Vardi~\cite{vardi1989multiplicative} studied the problem of estimating the 
distribution function of $X$
under the assumption that two samples $Y_1,\ldots,Y_n$ and $X_{n+1},\ldots, X_{n+m}$ are available.
The aforementioned paper develops a nonparametric maximum likelihood estimator; 
large sample properties of this estimator 
are studied in Vardi \& Zhang~\cite{vardi1992large}.
The problem of density estimation in the multiplicative censoring model was considered in  
Andersen \& Hansen~\cite{andersen2001multiplicative}
and Comte \& Dion~\cite{comte2016}, where estimators 
based on orthogonal  series have been developed.
Kernel density estimators were studied
in Asgharian et al.~\cite{asgharian2012large} and Brunel et al.~\cite{brunel2016nonparametric}.
We also refer the reader to the recent work  by Belomestny et al.~\cite{belomestny2016nonparametric}
where
a generalized multiplicative censoring model with 
$(\eta_i)$ being beta-distributed random variables was 
introduced and studied; see also references therein. 
\par
Third, as mentioned above, the outlined problem can be viewed as 
the problem of demixing of a scale mixture.
Closely related problems of estimating mixing densities 
were considered by Zhang~\cite{zhang1990fourier},~\cite{zhang1995onestimating} and 
Loh \& Zhang~\cite{loh1997estimating}.
In particular, the paper \cite{zhang1990fourier} 
develops Fourier techniques for estimating mixing densities in
location models, 
while \cite{zhang1995onestimating} and \cite{loh1997estimating}
focus on estimating mixing densities in discrete exponential family models.
However we are not aware of works on estimating mixing densities in the context of  scale models.
Finally, we also mention   related results on  estimating
regression functions with multiplicative errors--in--variables that are
reported in
Iturria et al.~\cite{iturria1999polynomial}.
\par 
A naive approach to the 
problem of density estimation in the model with multiplicative errors 
is based on 
reduction to the additive measurement error 
model. In particular, assuming that $X_i$'s and $\eta_i$'s are positive random variables and taking 
logarithms of the both sides of (\ref{eq:observations}), 
we come to the additive model $Y_i^\prime=X^\prime_i + \eta^\prime_i$, 
where $Y_i^\prime=\ln Y_i$, $X^\prime_i=\ln X_i$ and $\eta^\prime_i=\ln \eta_i$.  In this model, 
the density  $f_{X^\prime}$ of 
$X^\prime$ can be estimated using the well developed 
methodology for additive deconvolution problems
(see, e.g.,  \cite{zhang1990fourier} and \cite{fan1991ontheoptimal}),
and then an estimator for $f_X$ can be  obtained 
using the inverse transformation $f_X(x)=(1/x)f_{X^\prime}(\ln x)$.
This idea has been utilized in  Van Es \& Spreij~\cite{van2011estimation} and 
 Van Es et al.~\cite{van2003nonparametric}. 
 However, several questions about applicability of this approach arise. First, it can be used only if 
$X$ and $\eta$ are  nonnegative random variables. Second, it does not provide 
an estimator of $f_X$
at the origin $x=0$ since the inverse transformation is not well--defined there. 
Third, even if this approach is applicable, 
it is not clear 
whether the resulting estimator possesses  the desired optimality properties. 
\par 
In contrast to  voluminous literature on density deconvolution in the model with 
additive measurement errors, the problem of density estimation
from observations with multiplicative errors 
was studied to a much lesser extent.
In fact, it
was considered  
only for specific distributions of errors $(\eta_i)$ such as 
normal, uniform 
or beta, and 
the estimators proposed in the literature are tailored to a specific form of the error density $g$. 
In this context the following 
natural questions arise. 
How to estimate $f_X$ under general assumptions on the error density $g$? Which properties  
of the error density $g$ do affect the estimation accuracy, and 
what is the achievable accuracy 
in estimating $f_X$? What can be said about properties of the 
 deconvolution estimators based on the  logarithmic transformation of the data? 
 \par
The main goal of the present paper is  
to develop optimal  estimators of $f_X$ in a principled way 
under general assumptions on the error density $g$ and to  provide answers to the questions raised above.
 Our approach 
makes use of the Mellin transform which, in view of its properties, 
is an appropriate tool for constructing estimators in this setting. 
\par
We adopt minimax framework for measuring estimation accuracy. Specifically,  
accuracy of an estimator 
$\hat{f}_X(x_0)$ of $f_X(x_0)$ is measured by the maximal risk
\[
 \cR_n[\hat{f}_X; \Sigma]:= \sup_{f_X\in \Sigma} \Big[\bE_{f_X} |\hat{f}_X(x_0)-f_X(x_0)|^2\Big]^{1/2},
\]
where $\Sigma$ is a class of densities. Here and in what follows, 
$\bE_{f_X}$ denotes the expectation with respect 
to the distribution of the observations $Y_1,\ldots,Y_n$ 
when the unknown density of $X$ is $f_X$. The minimax risk
is defined by
\[
 \cR_n^*[\Sigma] := \inf_{\hat{f}_X} \cR_n[\hat{f}_X;\Sigma] = 
 \inf_{\hat{f}_X} \sup_{f_X\in \Sigma} \Big[\bE_{f_X} |\hat{f}_X(x_0)-f_X(x_0)|^2\Big]^{1/2},
\]
where $\inf$ is taken over all possible estimators.
Our goal is to develop an estimator $\hat{f}_X(x_0)$ which is {\em rate--optimal}, i.e., 
\[
 \cR_n[\hat{f}_X; \Sigma] \leq C_n \cR_n^*[\Sigma],\;\;\;\sup_n C_n <\infty.
\]
\paragraph{Main contributions}
The main contributions of this work are as follows. 
\par 
We elucidate the main feature of the multiplicative measurement errors 
setting: the influence of the estimation point $x_0$ on the achievable estimation accuracy.
In particular, assuming that unknown density $f_X$ belongs 
to a local H\"older functional class in a vicinity of $x_0$, we show that, depending on the value of $x_0$, 
there are 
two different regimes in terms of the rates of convergence of the minimax risk. We develop a general 
method for 
estimating $f_X(x_0)$ in these two regimes.  
\par
The first regime corresponds to the situation  when  the value of $x_0$ is separated away
from zero. Here the achievable rate of convergence 
is primarily 
determined by the value of $x_0$, by the local smoothness of $f_X$,
and by the 
ill--posedness of the integral transform in 
(\ref{eq:f_Y}). The latter is characterized in terms of the 
rate at which the Mellin transform of $g$ decreases at infinity on a line parallel to the imaginary 
axis in the complex plane.
It is worth noting that this characteristic is global 
in the sense that it is determined by the global behavior of the error density $g$ on its support.
We construct a 
kernel--type estimator of $f_X(x_0)$  and prove that it is rate--optimal
in terms of
dependence on the sample size $n$, parameters of the considered functional class $\Sigma$ and $x_0$.
It turns out that the deconvolution estimator based 
on the logarithmic transformation of the data is a 
special case of the proposed estimation procedure.
As a by--product of our general results, we demonstrate that if $x_0$ is separated away from zero, 
the random variables $X$ and $\eta$ are nonnegative, and 
 $f_X$ belongs to a local H\"older class  in a vicinity of $x_0,$ 
then under certain conditions on $g$ the deconvolution estimator is rate--optimal. 
However, if $f_X$ satisfies some additional constraints, e.g., a moment condition, 
then the accuracy of the deconvolution estimator can be improved. 
\par
In the second regime, where $x_0=0$,  completely different phenomena 
are observed. 
It turns out that in this case the achievable accuracy in estimating $f_X(0)$ 
is determined by smoothness of $f_X$ and by  local behavior of $g$
in  vicinity of the origin. Thus, in contrast to the first regime, 
 the minimax 
rate depends only on  local characteristics of $g$ and is not affected 
by the ill--posedness of the integral transform in 
(\ref{eq:f_Y}). 
In particular, our results imply that if $g$ is bounded 
and does not vanish in a vicinity of the origin, then 
the minimax rate of convergence  
is only  by a  $\ln n$--factor
worse than the one achievable in the problem of density estimation from direct observations. 
We also construct a rate--optimal 
estimator of $f_X(0)$ and prove a matching lower bound on the minimax risk. 
\paragraph{Organization of the paper} 
The rest of the paper is organized as follows.  
In Section~\ref{sec:prelim} we introduce notation,
discuss some properties of the Mellin transform that are used throughout the paper and present 
an identifiability result.
Section~\ref{sec:nonzero} deals with the setting when $x_0$ is separated away from zero; we construct
estimators under different assumptions on the error density $g$ 
and present results on their accuracy over suitable classes of densities.
Section~\ref{sec:zero} is devoted to the problem of estimating \(f_X(0)\). 
A  simulation study of the proposed estimators is 
presented in Section~\ref{sec:sim}. 
Finally,  proofs of  main results are presented in Section~\ref{sec:proofs} while 
proofs of auxiliary statements are given in
Section~\ref{sec:auxiliary}.

\section{Preliminaries}
\label{sec:prelim}
In this section we introduce notation and discuss basic properties of the Mellin transform that 
will be extensively used throughout the paper. This material can be found, e.g., in \cite{paris2001asymptotics}
and \cite{wong2001asymptotic}.
In addition, we present a result on 
identifiability of the distribution of $X$ in the model~(\ref{eq:observations}).
\paragraph{The Mellin transform}
For a generic locally integrable function $u$ on $(0,\infty)$ 
the Mellin transform of $u$ is defined by
\begin{equation}\label{eq:M}
 \widetilde{u}(z)= \cM[u;z]:=\int_0^\infty x^{z-1} u(x) \rd x
\end{equation}
for all $z\in \bC$ such that the integral on the right hand side 
is absolutely convergent. 
The region of convergence $\Omega_u$ 
is an infinite 
vertical strip in the complex plane $\bC$,
\[
\Omega_u=\{z\in \bC: a<{\rm Re}(z)<b\}, \;\;a<b, 
\]
or a vertical line $\Omega_u=\{z: {\rm Re}(z)=c\}$ if 
$u(x) x^{c-1} \in \bL_1(\bR_+)$ for one $c\in \bR$.
For example, if $u(x)=O(x^{-a+\epsilon})$ as $x\to 0+$ and $u(x)=O(x^{-b-\epsilon})$ as $x\to\infty$ 
for some $\epsilon>0,$ then the integral in (\ref{eq:M}) converges absolutely and defines an analytic function 
$\widetilde{u}(z)$ on $\Omega_u=\{z: a<{\rm Re}(z)<b\}$.
\par
The inversion formula for the Mellin transform is 
\[
 u(x)=\frac{1}{2\pi i}\int_{c-i\infty}^{c+i\infty} x^{-z} \widetilde{u}(z)\,\rd z,\;\;\;c\in \Omega_u\cap (-\infty, \infty). 
\]
\par
Let $u(x)$ and $v(x)$ be functions  such that the integral 
$I = \int_0^\infty u(x)v(x)\rd x$ 
exists.
Assume also that the Mellin transforms $\widetilde{u}(1-z)=\cM[u;1-z]$ and 
$\widetilde{v}(z)=\cM[v;z]$ have a common strip of analyticity, 
which will be the case when $I$ is absolutely convergent. Then for any line $\{z:\,{\rm Re}(z)=c\}$ in this
common strip the Parseval formula is valid:
\[
 \int_0^\infty u(x)v(x)\rd x = \frac{1}{2\pi i}\int_{c-i\infty}^{c+i\infty} \widetilde{u}(1-z)
 \widetilde{v}(z)\rd z.
\]
In particular, we get for $u=v$ and $c=\frac{1}{2}$,
\[
 \int_0^\infty u^2(x) \rd x = \frac{1}{2\pi}\int_{-\infty}^\infty |\widetilde{u}(\tfrac{1}{2}+i\omega)|^2 \rd 
 \omega.
\]
It also holds  
\begin{equation}\label{eq:parceval-1}
\int_0^\infty u^2(x) x^{2s-1}\rd x = \frac{1}{2\pi}\int_{-\infty}^\infty 
|\widetilde{u}(s+i\omega)|^2\rd \omega.     
    \end{equation} 
\par 
Let us mention the relation   
of the Mellin transform to a multiplicative convolution integral \eqref{eq:f_Y}; this property is  
central in subsequent developments. Let $u$ and $v$ be defined on $[0,\infty)$, and let
\[
 [u\star v](y):=\int_0^\infty \frac{1}{x} u(x) v(y/x) \rd x;
\]
then 
\[
\widetilde{[u\star v]}(z) = \cM[u\star v; z]=\cM[u; z]\cM[v; z]=\widetilde{u}(z)\widetilde{v}(z).
\]
\par
We shall use the Mellin transform techniques  
for functions defined on the whole real line. 
To this end, for  a function $u$ on $(-\infty, \infty)$ we  set
\begin{equation}\label{eq:u+u-}
u^+(x):=\left\{\begin{array}{ll}
                u(x), & x\geq 0,\\
                0, & x<0
               \end{array}
               \right.
\;\;\;\;\hbox{and}\;\;\;\;\;u^-(x):=\left\{\begin{array}{ll}
                                       u(-x), & x> 0,\\
                                       0, & x\leq 0.
                                      \end{array}
\right.
\end{equation}
It is evident that 
with this notation $u(x)=u^+(x)$ for $x\geq 0$ and $u(x)=u^{-}(-x)$ for $x<0$.
%
The one--sided Mellin transforms of 
function $u$ defined on $(-\infty, \infty)$ are given~by
\begin{eqnarray*} 
&&\widetilde{u}^+(z)=\int_0^\infty x^{z-1}u^+(x)\rd x = \int_0^\infty x^{z-1}u(x)\rd x,
\\
&&\widetilde{u}^-(z)=\int_0^\infty x^{z-1}u^-(x)\rd x = \int_{-\infty}^0 (-x)^{z-1}u(x)\rd x.
\end{eqnarray*}
\paragraph{The Laplace and Fourier transforms} 
The bilateral Laplace transform of function $u$ on $(-\infty,\infty)$ is defined as
\[
\widecheck{u}(z) = \cL[u; z]:=\int_{-\infty}^\infty u(x) e^{-zx}\rd x,
\]
and if the integral absolutely converges on a line $\{z:{\rm Re}(z)=c\},$ then 
the  inverse Laplace
transform is given by
\[
 u(x)=\frac{1}{2\pi i} \int_{c-i\infty}^{c+i\infty} \widecheck{u}(z)e^{zx} \rd z.
\]
The Fourier transform of $u$ is 
$ \widehat{u}(\omega)=\cF[u;\omega]:=\cL[u; i\omega] =\widecheck{u}(i\omega)$.
\paragraph{Identifiability} 
In the model (\ref{eq:observations}) we do not assume that 
the random variables $X$ and $\eta$ are nonnegative. 
This fact raises the question whether 
the distribution of $X$ is identifiable from 
the distribution of $Y$. The next statement provides a necessary and sufficient 
condition for the identifiability.  
\begin{lemma}\label{lem:identifiability}
The probability density $f_X$ is identifiable  from $f_Y$ 
if and only if  $g(x)\ne g(-x)$ on a set of positive Lebesgue measure.
\end{lemma}
The proof of Lemma~\ref{lem:identifiability} 
is given in Section~\ref{sec:auxiliary}. It 
shows that the identifiability condition is equivalent to the requirement 
that $|[\widetilde{g}^+(z)]^2-[\widetilde{g}^-(z)]^2|$ is not zero for almost all $z$ in the common strip of analyticity 
of $\widetilde{g}^+$ and $\widetilde{g}^-$.
Finally, we note that if one of the variables $X$ or $\eta$ is 
nonnegative, then the condition of identifiability is trivially fulfilled.

\section{Estimation at a point separated away from zero}
\label{sec:nonzero}
In this section we consider 
the problem of  estimation of \(f_X\) at a point \(x_0\) separated away from zero. 
\par
\subsection{Construction of estimator}
We adopt the  linear functional strategy 
for constructing our estimators. This strategy has been 
frequently 
used
for solving ill--posed inverse problems
(see, e.g., \cite{goldberg1979amethod} and \cite{anderssen1980ontheuse}).
In our context, the main idea of this 
method is to find a pair of kernels, say, $K(x,y)$ and $L(x,y)$ 
such that: 
\begin{itemize}
\item[(i)] $\int_{-\infty}^\infty K(x,y) f_X(y)\rd y$ 
approximates ``well'' the value  $f_X(x)$ 
to be recovered; 
\item[(ii)] kernel $L(x,y)$ is related to $K(x,y)$ via the equation
\begin{equation}\label{eq:lin-func}
 \int_{-\infty}^\infty K(x,y) f_X(y)\rd y = \int_{-\infty}^\infty L(x,y) f_Y(y)\rd y.
\end{equation}
\end{itemize}
Then under (i) and (ii),
the empirical estimator of the integral on the right hand side 
of (\ref{eq:lin-func}) provides  a sensible estimator for $f_X(x)$.  
\paragraph{Kernel construction}
Let
$K:\bR\to \bR$ be a kernel function
and for any positive real number $h$ define 
\begin{equation}\label{eq:K-kernel}
 K_{h}(x, y)=\left\{\begin{array}{ll} \tfrac{1}{x h} K\big(\tfrac{\ln (y/x)}{h}\big), & y/x>0,
 \\*[2mm]
              0, & y/x<0.
             \end{array}
             \right.
\end{equation}
Let  
$\widetilde{g}^+(z)=\cM[g^+; z]$ 
and $\widetilde{g}^-(z)=\cM[g^-;z]$
be the one--sided Mellin transforms of $g$, 
and let 
\begin{equation}\label{eq:strip}
\Omega_{g^+} \cap \Omega_{g^-} = : \{z\in \bC: a<{\rm Re}(z)<b\}
\end{equation}
be the common strip of their analyticity. 
Since $g$ is a probability density, we always have  $a<1<b$; hence   
$\Omega_{g^+}\cap \Omega_{g^-}$ is non--empty -- it always contains the line $\{z\in \bC: {\rm Re}(z)=1\}$. 
We note that 
$\Omega_{g^+}$ and/or $\Omega_{g^-}$ can degenerate to this line. In this case, 
by convention, we put $a=1$, $b=1$, and corresponding open interval  should be replaced
by a singleton.
\par 
For $s\in (1-b, 1-a)$ define
\begin{multline}\label{eq:L}
  L_{s,h}(x,y) 
  \\
  := \left\{ 
  \begin{array}{ll}
  {\displaystyle 
  \frac{1}{2\pi i x} \int_{s-i\infty}^{s+i\infty} 
  \Big|\frac{x}{y}\Big|^z\,  \frac{\widecheck{K}(zh)\;\widetilde{g}^+(1-z)}{[\widetilde{g}^+(1-z)]^2 - [\widetilde{g}^-(1-z)]^2}
  \,\rd z,} & y/x>0,
  \\*[6mm]
  {\displaystyle
   -\frac{1}{2\pi i x} \int_{s-i\infty}^{s+i\infty} 
   \Big|\frac{x}{y}\Big|^z  \frac{\widecheck{K}(zh)\; \widetilde{g}^-(1-z)}{[\widetilde{g}^+(1-z)]^2 - [\widetilde{g}^-(1-z)]^2}
  \,\rd z,}  & y/x<0.
  \end{array}
  \right.
\end{multline}
For the time being, we suppose that 
the kernel $K$ and the error density $g$ are such that the function $L_{s,h}$ is well defined; 
the corresponding conditions  
on $K$ and $g$ will be formulated later. 
Several remarks on this definition are in order. 
\begin{remark}
\begin{itemize}
\item[{\rm (i)}] We can assume that the Laplace transform $\widecheck{K}(\cdot)$
of kernel $K$ is an entire function.
This does not restrict generality since $K$ can be 
always chosen to satisfy this assumption.  
\item[{\rm (ii)}]
If $[\widetilde{g}^+(z)]^2- [\widetilde{g}^-(z)]^2 \neq 0$ for all 
$z\in \Omega_{g^+}\cap \Omega_{g^-}$ then the integrands in (\ref{eq:L}) are  analytic functions in 
$\{z\in \bC: 1-b < {\rm Re}(z) <1-a\}$. In this case the integrals in (\ref{eq:L}) do not depend on the integration
path, and $L_{s,h}(x,y)$ does not depend on $s\in (1-b, 1-a)$. If 
function $[\widetilde{g}^+(z)]^2- [\widetilde{g}^-(z)]^2$ has zeros in $\Omega_{g^+}\cap \Omega_{g^-}$ 
then the functions under the integral  sign in 
(\ref{eq:L}) are meromorphic, and $L_{s,h}(x,y)$ depends on parameter $s$.
\end{itemize}
\end{remark}
\par 
The relationship between kernels $L_{s,h}(x,y)$ and $K_h(x,y)$ 
in \eqref{eq:L} and \eqref{eq:K-kernel} is revealed in  the following statement.
\begin{lemma}\label{lem:L}
Let $K_h(x,y)$ be given by (\ref{eq:K-kernel}).
Let $s\in (1-b, 1-a)$ where $a$ and $b$ are given in (\ref{eq:strip}), and suppose that
the integrals on the right hand side of  (\ref{eq:L}) are absolutely convergent.
Then  
it holds that 
\begin{equation}\label{eq:K-L}
 \int_{-\infty}^\infty L_{s,h}(x,y) f_Y(y) \rd y = \int_{-\infty}^\infty K_h(x,t) f_X(t)\rd t.
\end{equation}
\end{lemma}
The proof of Lemma~\ref{lem:L} is given in Section~\ref{sec:auxiliary}. 
We note that relationship (\ref{eq:K-L}) is 
in full accordance with  the linear functional strategy [cf.~(\ref{eq:lin-func})].
Because $a<1<b$, it holds that $0\in (1-b, 1-a)$; 
 hence one can always choose $s=0$ in (\ref{eq:L}). This choice yields
 \begin{eqnarray*}
 L_{0,h}(x,y) = \left\{ 
  \begin{array}{ll}
  {\displaystyle 
  \frac{1}{2\pi x} \int_{-\infty}^{\infty} 
  \Big|\frac{x}{y}\Big|^{i\omega}\,  \frac{\widehat{K}(\omega h)\;\widetilde{g}^+(1-i\omega)}
  {[\widetilde{g}^+(1-i\omega)]^2 - [\widetilde{g}^-(1-i\omega)]^2}
  \,\rd \omega,} & y/x>0,
  \\*[6mm]
  {\displaystyle
   -\frac{1}{2\pi  x} \int_{-\infty}^{\infty} 
   \Big|\frac{x}{y}\Big|^{i\omega}  
   \frac{\widehat{K}(\omega h)\; \widetilde{g}^-(1-i\omega)}{[\widetilde{g}^+(1-i\omega)]^2 - 
   [\widetilde{g}^-(1-i\omega)]^2}
  \,\rd \omega,}  & y/x<0.
  \end{array}
  \right.
\end{eqnarray*}
If $g$ is supported on $[0,\infty)$, 
then $\widetilde{g}^-=0$, $\widetilde{g}^+=\widetilde{g}$;  in this case  
 \begin{eqnarray}\label{eq:L-1}
  L_{s,h}(x,y) =
  \frac{1}{2\pi i x} \int_{s-i\infty}^{s+i\infty} 
  \Big|\frac{x}{y}\Big|^z\,  \frac{\widecheck{K}(zh)}{\widetilde{g}(1-z)}
  \,\rd z, \;\;\;y/x>0,
\end{eqnarray}
and $L_{s,h}(x,y)=0$ whenever $x/y<0$.
In particular, 
for $s=0$ we have 
\begin{equation}\label{eq:L-e>0}
 L_h(x,y) := L_{0,h}(x,y)= \frac{1}{2\pi x}\int_{-\infty}^\infty \Big|\frac{x}{y}\Big|^{i\omega}\frac{\widehat{K}(\omega h)}
 {\widetilde{g}(1-i\omega)}\rd \omega, \;\;\;y/x>0.
\end{equation}
\paragraph{Estimator}  
For $|x_0|>0$ we define the estimator of $f_X(x_0)$ by
\begin{equation}\label{eq:estimator}
 \hat{f}_{s,h}(x_0)=\frac{1}{n}\sum_{j=1}^n L_{s,h}(x_0, Y_j),
\end{equation}
where 
$L_{s,h}$ is given in (\ref{eq:L}), 
$h>0$ and  $s\in (1-b, 1-a)$ are two tuning parameters to be specified.
In what follows with a slight abuse of notation we shall write $\hat{f}_h(x_0):=\hat{f}_{0,h}(x_0)$
and  $L_h(x,y):=L_{0,h}(x,y)$.
\par 
Note also that (\ref{eq:K-L}) implies 
\[
 \bE_{f_X} [\hat{f}_{s,h}(x_0)]= \int_{-\infty}^\infty K_h(x_0, t) f_{X}(t)\rd t.
\]
The latter formula  is 
crucial for the analysis of the bias of \(\hat{f}_{s,h}(x_0).\)  

\subsection{Relation to the additive deconvolution problem}\label{sec:relation}  
There is close   
 connection between the kernel $L_h(x,y)=L_{0,h}(x,y)$ defined in  (\ref{eq:L-e>0}) 
 and  kernels used in the additive 
 deconvolution problems. 
Specifically, suppose that $X$ and $\eta$ are positive random variables, and let 
 $\eta^\prime=\ln \eta$. If $g$ is the density of $\eta$, and $\widehat{g}$ is 
 the corresponding characteristic function, then $g_{\eta^\prime}(x)=e^x g(e^x)$ is
 the density of $\eta^\prime$, and 
 the characteristic function of $\eta^\prime$
 is $\widehat{g}_{\eta^\prime}(\omega)=\cM[g; 1-i\omega]=\widetilde{g}(1-i\omega)$.
Therefore the expression for $L_h(x,y)$ in (\ref{eq:L-e>0}) can be rewritten as
\[
 L_h(x,y)=\frac{1}{2\pi x}\int_{-\infty}^\infty 
 \frac{\widehat{K}(\omega h)}{\widehat{g}_{\eta^\prime}(\omega)}
 e^{-i\omega (\ln y-\ln x)} \rd \omega,\;\;\;x>0, \;y>0,
\]
and 
the corresponding estimator of $f_X(x_0)$ [cf. (\ref{eq:estimator})] is 
\begin{equation}\label{eq:mult-dec-est}
 \hat{f}_X(x_0) = \frac{1}{n}\sum_{j=1}^n L_h(x_0, Y_j)=
 \frac{1}{2\pi x_0 n}\sum_{j=1}^n \int_{-\infty}^\infty
 \frac{\widehat{K}(\omega h)}{\widehat{g}_{\eta^\prime}(\omega)}
 e^{-i\omega (\ln Y_j -\ln x_0)} \rd \omega.
\end{equation}
\par 
On the other hand, consider  the additive deconvolution model for the logarithms, 
$Y^\prime=X^\prime+\eta^\prime$, 
 where $Y^\prime=\ln Y$, $X^\prime=\ln X$ and $\eta^\prime=\ln \eta$. 
Then the standard deconvolution estimator
of $f_{X^\prime}(t_0)$ is of the form
\begin{equation*}
 \hat{f}_{X^\prime}(t_0)=\frac{1}{2\pi n}\sum_{j=1}^n
 \int_{-\infty}^\infty \frac{\widehat{K}(\omega h)}{\widehat{g}_{\eta^\prime}(\omega)} 
 e^{-i\omega (Y_j^\prime - t_0)}\rd \omega.
\end{equation*}
Since $f_{X^\prime}(t_0)=e^{t_0}f_X(e^{t_0})$, we can estimate 
$f_X(x_0)=\frac{1}{x_0}f_{X^\prime}(\ln x_0)$ by 
\begin{eqnarray}
\hat{f}_X(x_0)
&=&\frac{1}{x_0}\hat{f}_{X^\prime}(\ln x_0)
\nonumber
\\
&=&\frac{1}{2\pi x_0 n}\sum_{j=1}^n
 \int_{-\infty}^\infty \frac{\widehat{K}(\omega h)}{\widehat{g}_{\eta^\prime}(\omega)} 
 e^{-i\omega (Y_j^\prime - \ln x_0)}\rd \omega
 \label{eq:add_log_est}
\end{eqnarray}
which coincides with (\ref{eq:mult-dec-est}).
\par 
We conclude that
if random variables $X$ and $\eta$ are positive, and the parameter $s$ 
of the estimator $\hat{f}_{s, h}(x_0)$ in (\ref{eq:estimator}) 
is set to zero, then both approaches lead to the same estimator. 
Thus,  the estimator \eqref{eq:add_log_est} is a particular case 
of our estimator $\hat{f}_{s,h}(x_0)$ defined in (\ref{eq:estimator}). 
We note however that tuning parameter $s$ adds some flexibility, and 
its  proper choice  
can  improve  
accuracy of  \(\hat{f}_{s,h}(x_0)\) under suitable assumptions (see, e.g., Theorem~\ref{th:upper-bound-2} below).
%
 
\subsection{Convergence analysis} \label{sec:analysis}
We proceed with convergence analysis of the risk of the proposed estimator $\hat{f}_{s,h}(x_0)$. 
In order to avoid unnecessary  
technicalities,   from now on we will assume that 
$X$ and $\eta$ are nonnegative random variables,~i.e.,
\begin{equation}\label{eq:simple}
{\rm supp}(g)\subseteq [0,\infty), \;
\Omega_g=\{z\in \bC: a<{\rm Re}(z)<b\},\;{\rm supp}(f_X)\subseteq [0,\infty)
\end{equation}
for some \(a>0\) and \(b>a\). Under these conditions the kernel 
$L_{s,h}(x, y)$ is  given by  (\ref{eq:L-1}).
\par
Assumption (\ref{eq:simple})  
streamlines the presentation and,  in fact,
does not lead to loss of generality. In particular, 
the ensuing analysis of the risk of $\hat{f}_{s,h}(x_0)$
remains valid for general random variables $X$ and $\eta$, provided that the conditions imposed in the sequel on the Mellin 
transform $\widetilde{g}$ of $g$
are replaced by the corresponding conditions on 
$([\widetilde{g}^+]^2-[\widetilde{g}^-]^2)/\widetilde{g}^+$ and 
$([\widetilde{g}^+]^2-[\widetilde{g}^-]^2)/\widetilde{g}^-$ [cf.~(\ref{eq:L})].
\par 
The risk of $\hat{f}_{s,h}(x_0)$ will be analyzed under a local smoothness assumption on $f_X$ and   
two different sets of assumptions on the error density~$g$.
\begin{definition} 
 Let $\beta>0$, $A>0$, $x_0>0$ and $r>1$.
We say that $f\in \sH_{x_0,r} (A,\beta)$  if 
$f$ is a probability density, that is, 
$\ell =\lfloor \beta\rfloor :=\max\{k\in \bN_0: k<\beta\}$ times continuously differentiable, 
and $\max_{k=1,\ldots, \ell} |f^{(k)}(x)|\leq A$,
\[
 \big|f^{(\ell)}(x)-f^{(\ell)}(x^\prime)\big| \leq 
 A|x-x^\prime|^{\beta -\ell},\;
 \forall x, x^\prime \in [r^{-1}x_0, rx_0].
\]
\end{definition}
\par 
As for the conditions on the error density $g$, some assumptions  characterizing
the rate of decay of the Mellin transform 
$\widetilde{g}(\sigma+i\omega)$ as $|\omega|\to\infty$ 
for a fixed $\sigma\in \Omega_g$ will be considered.
Depending on the tail behavior of $\widetilde{g},$ we distinguish between the following two cases:
\begin{itemize}
 \item {\em smooth error densities}, when the tails of $\widetilde{g}$ are polynomial, i.e.,
 \[
  \widetilde{g}(\sigma+i\omega) \asymp |\omega|^{-\gamma},\;\;|\omega|\to \infty, \;\;\sigma\in \Omega_g
 \]
\item {\em super--smooth error densities}, when the tails of $\widetilde{g}$ are exponential, i.e.,
\[
 \widetilde{g}(\sigma+i\omega) \asymp \exp\{-\gamma |\omega|\},\;\;|\omega|\to \infty,\;\;\;\sigma\in \Omega_g.
\]
\end{itemize}
Our terminology here 
is similar to that used
in the additive deconvolution problem,
even though the words {\em smooth} and {\em super--smooth} should not be understood literally.
\subsubsection{Smooth error densities}
The class of smooth error densities is determined by the following assumption.
\begin{itemize}
 \item[\hbox{[G1]}] 
 For some $\sigma\in  (a, b),$ 
 there exist real numbers $\omega_0>0$, $c_0>0$, $B_2>B_1>0$ and $\gamma>0$ 
 such~that 
 \begin{align}
 & \;\min_{|\omega|\leq \omega_0} |\widetilde{g}(\sigma+i\omega)|\geq c_0 >0,\;
 \nonumber
 \\*[2mm]
 & B_1 |\omega|^{-\gamma} \leq |\widetilde{g}(\sigma + i\omega)| \leq B_2 |\omega|^{-\gamma},\;\;
  \forall |\omega|\geq \omega_0. 
 \label{eq:as1-2}
 \end{align}
\end{itemize}
\par 
We will require Assumption [G1] for a particular choice of 
$\sigma\in (a,b)$, and  parameters $c_0$, $\omega_0$, $B_1$, $B_2$ and $\gamma$
may depend on $\sigma$.
Assumption~[G1] stipulates the rate of decay of $\widetilde{g}$ on the line 
$\{z: {\rm Re}(z)=\sigma\}$ as $|{\rm Im}(z)|\to \infty$ and 
implies that $\tilde{g}$ does not have zeros on this line.
This requirement  
is similar to standard assumptions 
in the additive deconvolution problem
on the rate of decay of the error characteristic function. 
 The following examples 
 show that [G1] 
 holds for many well-known distributions.
\begin{example}[a Beta distribution] \label{ex:1}
Let $g(x)=(\nu+1)x^\nu/\theta^{\nu+1}$, $0<x<\theta$ with $\nu>-1$; 
 then 
 \[
 \widetilde{g}(z)=(\nu+1)\theta^{z-1}/(\nu+z),\quad {\rm Re}(z)>-\nu,
 \]
$a=-\nu$, $b=\infty$, and
\[
 |\widetilde{g}(\sigma+i\omega)|=\theta^{\sigma-1} (\nu+1)[(\nu+\sigma)^2+\omega^2]^{-1/2}, \quad\sigma>-\nu.
\]
Then Assumption~[G1] is verified for any $\sigma>-\nu$ with 
$\gamma=1$, $\omega_0= 2(\sigma+\nu)$, \(c_0=(1/5)^{1/2}\theta^{\sigma-1} (\nu+1)/(\nu+\sigma)\) and 
$B_1=(4/5)^{1/2}\theta^{\sigma-1}(\nu+1)$, $B_2=\theta^{\sigma-1}(\nu+1)$.
 The case $\nu=0$, $\theta=1$
 corresponds to the uniform distribution with
 $\widetilde{g}(z)=1/z$ and $|\widetilde{g}(\sigma+i\omega)|=(\sigma^2+\omega^2)^{-1/2}$ for 
 $\sigma>0$. 
\end{example}
\begin{example}[Pareto's distribution] \label{ex:2}
 Let $g(x)=(\nu-1) \theta^{\nu-1} /x^{\nu}$, $x>\theta$ with $\theta>0$ and $\nu>1$. Then  
\[
 \widetilde{g}(z)=(\nu-1) \theta^{z-1}/(\nu-z),\quad  {\rm Re}(z)<\nu,
\]
$a=-\infty$, $b=\nu$, and 
\[
 |\widetilde{g}(\sigma+i\omega)| = 
 (\nu-1) \theta^{\sigma-1}[(\nu-\sigma)^2+\omega^2]^{-1/2},\quad \sigma<\nu.
\]
Hence
Assumption~[G1] is verified for any $\sigma<\nu$ with $\gamma=1$, 
$\omega_0= 2(\nu-\sigma)$, \(c_0=(1/5)^{1/2} (\nu-1) \theta^{\sigma-1}/(\nu-\sigma)\),
$B_1=(4/5)^{1/2}\,(\nu-1)\theta^{\sigma-1}$, $B_2=(\nu-1)\theta^{\sigma-1}$.
\end{example}
\begin{example} \label{ex:3} Natural 
examples of random variables whose distributions satisfy Assumption~[G1] with $\gamma>1$ 
can be obtained
by multiplication of independent random variables with densities as in Examples~\ref{ex:1} and~\ref{ex:2}.
 For instance, 
 the probability density of a random variable which is a product of two independent random variables
 uniformly distributed on $[0,1]$ is $g(x)=\ln (1/x)$, $0\leq x\leq 1$. 
 For this density $\widetilde{g}(z)=1/z^2$ and $|\widetilde{g}(\sigma+i\omega)|=(\sigma^2+\omega^2)^{-1}$,
 so that Assumption~[G1] holds with $\gamma=2$.
\end{example}

\paragraph{Bounds on the risk}
We begin with establishing an upper bound 
on the risk of the estimator $\hat{f}_{s,h}(x_0)$ 
under Assumption~[G1].
\par 
In this case the kernel $K$ is chosen to satisfy the following conditions.  
Assume that
$K:\bR\to\bR$ is a bounded function that vanishes outside $[-1,1]$ and satisfies
\begin{itemize}
\item[(i)] for  a positive integer number $m,$
\begin{eqnarray}\label{eq:kernel-K}
&&
\int_{-1}^1 K(t)\rd t=1,\;\;
\int_{-1}^1 t^kK(t)\rd t=0,\quad  k=1,\ldots, m;
\end{eqnarray}
\item[(ii)] for a positive integer number $q,$ function $K$ is $q$ times continuously differentiable on $\bR$ and
for $j=0,1,\ldots, q$
\begin{equation}\label{eq:kernel-K-2}
\max_{x\in [-1,1]} |K^{(j)}(x)|
\leq C_{K}<\infty.
\end{equation}
\end{itemize}
%
%
\begin{theorem}\label{th:upper-bound-G1} Fix some \(\beta>0,\) \(r>0,\) \(A>0,\)  $x_0>0$ and consider the class $\sH_{x_0,r} (A,\beta).$  Suppose that  
Assumption~[G1] holds with $\sigma=1$ and some $\gamma>1$. Let 
$\hat{f}_{h_*}(x_0)=\hat{f}_{0,h_*}(x_0)$ be the estimator defined in
(\ref{eq:L-e>0})--(\ref{eq:estimator}) and associated with a kernel $K$ satisfying 
(\ref{eq:kernel-K})--(\ref{eq:kernel-K-2}) with parameters
$m\geq \lfloor \beta\rfloor +1$, $q> \gamma+1$, and 
\begin{equation}\label{eq:h-*}
 h=h_*:= \big[A^2 x_0^2(x_0^\beta+1)^2 n\big]^{-\frac{1}{2\beta+2\gamma+1}}.
\end{equation}
Then for $h_*<\min\{\ln r,1\}$ it holds that
\begin{equation}\label{eq:R-1}
 \cR_n\big[\hat{f}_{0,h_*}; \sH_{x_0,r}(A,\beta)\big] 
 \;\leq\; C_1 
 \big[A(x_0^\beta+1)\big]^{\frac{2\gamma+1}{2\beta +2\gamma +1}} 
 \big(x_0^{2} n\big)^{-\frac{\beta}{2\beta+2\gamma+1}},
\end{equation}
where $C_1$ depends on $\beta$ only.
\end{theorem}
Several remarks on the result of Theorem~\ref{th:upper-bound-G1} are in order. 
\begin{remark}\label{rem:1}
\mbox{}
\begin{itemize}
\item[{\rm (i)}] If $\gamma\leq 1,$ then the result of 
Theorem~\ref{th:upper-bound-G1} holds for a slightly smaller
set of functions than $\sH_{x_0,r}(A,\beta)$. In particular, if 
\begin{equation}\label{eq:gamma<1}
f_X\in \sH_{x_0,r}(A,\beta) \cap \bigg\{f_X: 
\int_{-\infty}^\infty \frac{|\widetilde{f}_X(1+i\omega)|}{
(1+|\omega|)^{\gamma}} \rd \omega \;\leq\;c<\infty \bigg\},
\end{equation}
for some \(c>0,\) then $\widetilde{f}_Y(1+i\omega)$ is integrable, and the statement of Theorem~\ref{th:upper-bound-G1}
is still valid. Note that this additional condition on 
$\widetilde{f}_X$ is very mild: by the Riemann--Lebesgue
lemma $\widetilde{f}_X(1+i\omega)\to 0$ as $|\omega|\to\infty$. 
\item[{\rm (ii)}] 
The above upper bound  critically depends  on the value of $x_0$. 
If  $x_0$ is separated away from zero by a constant, then for large enough $n$ 
the bound takes the form  
\begin{equation}\label{eq:x0-large}
 \cR_n\big[\hat{f}_{h_*}; \sH_{x_0,r}(A,\beta)\big] 
 \leq C_2A^{\frac{2\gamma+1}{2\beta+2\gamma+1}} 
(x_0^{2\gamma-1}n^{-1})^{\frac{\beta}{2\beta+2\gamma+1}}.
\end{equation}
In particular, this shows that estimation accuracy gets worse for larger values of $x_0$.
\end{itemize}
\end{remark}
\par 
Now we establish a lower bound on the minimax risk under Assumption~[G1].  
We require the following additional condition on the error density~$g$.
\begin{itemize}
\item[\hbox{[G1$^\prime$]}]
For $\sigma \in (a,b)$ the first derivative of  $\widetilde{g}$ satisfies 
\[
 |\widetilde{g}^\prime (\sigma +i \omega)|\leq B |\omega|^{-\gamma},\;\;\;\forall |\omega|\geq \omega_0.
\]
\end{itemize}
\par 
Assumption~[G1$^\prime$] is similar to standard conditions
on derivatives of the characteristic function of the measurement error distribution
in the proofs of lower bounds for density deconvolution;
cf., e.g., Theorem~5 in \cite{fan1991ontheoptimal}. 
%
\begin{theorem}\label{th:lower-bound-G1}
Let  $x_0 \geq C_3>0$   
 for some constant $C_3$, 
and suppose that Assumptions~[G1] and~[G1$^\prime$] hold with $\sigma=1$ and $\gamma>1/2$.
Then 
\[
 \liminf_{n\to\infty} \;\Big\{\,\phi_n^{-1} \cR_n^*[\sH_{x_0,r}(A, \beta)]\,\Big\} \geq C_4, 
 \]
 where
 \[ \phi_n:= 
A^{\frac{2\gamma+1}{2\beta+2\gamma+1}} \big(x_0^{2\gamma-1}
n^{-1}\big)^{\frac{\beta}{2\beta+2\gamma+1}},
\]
and  
$C_4$ depends on $\beta$ and $r$ only.
\end{theorem}
\par
\begin{remark}\label{rem:3}
 \mbox{}
 \begin{itemize}
  \item[{\rm (i)}] Note that the lower bound of 
 Theorem~\ref{th:lower-bound-G1} 
 coincides  with the upper bound (\ref{eq:x0-large}) in terms of its dependence on $n$, $x_0$ and~$A$. 
 This implies  that for $x_0$ separated away from zero, 
 the estimator $\hat{f}_{h_*}(x_0)$
 is  rate--optimal, and  
 dependence  of the risk on $x_0$ over the functional class $\sH_{x_0,r}(A,\beta)$ cannot be improved. 
 \item[{\rm (ii)}]  In view of the interpretation  of $\hat{f}_{h_*}(x_0)$ given in  Section~\ref{sec:relation}, 
Theorems~\ref{th:upper-bound-G1} and~\ref{th:lower-bound-G1} assert 
rate--optimality of the standard deconvolution estimator in the additive measurement error model
based on   the log--transformed 
data, provided that the bandwidth parameter $h_*$ is selected as  in (\ref{eq:h-*}). 
Note  however that the standard choice  of \(h\) in additive deconvolution  does not involve \(x_0.\)  
\item[{\rm (iii)}] The proof of the lower bound in Theorem~\ref{th:lower-bound-G1} is based on the reduction
to a two--point hypotheses testing problem when under the null hypothesis
\[
 f_X(x)=f_X^{(0)}(x):=\frac{1}{\pi x(1+\ln^2(x/x_0))}, \;\;x>0. 
\]
The convergence region of 
the Mellin transform $\widetilde{f}_X^{(0)}(z)$ of $f_X^{(0)}(x)$ is the line 
$\{z: {\rm Re}(z)=1\}$, and this fact is essential
for the result of Theorem~\ref{th:lower-bound-G1}. 
If the Mellin transform is analytic in a non--degenerating strip 
around $\{z: {\rm Re}(z)=1\}$ then, under certain assumptions on measurement error density $g$, 
the estimation accuracy can be improved in terms of dependence on $x_0$. This issue is 
a subject of the next paragraph.
 \end{itemize}
\end{remark}
 \paragraph{Choice of parameter $s$ and improvements} 
 It is important to realize the interplay between conditions on $g$ and $f_X$ that lead to the results of 
Theorems~\ref{th:upper-bound-G1} and~\ref{th:lower-bound-G1}. 
In particular, the following two facts are essential for the stated results.
\begin{itemize}
 \item[(a)]  Since $f_X$ is a probability density, the Mellin transform $\widetilde{f}_X(z)$ always exists 
 on the vertical line 
 $\{z: {\rm Re}(z)=1\}$. 
 Note however that the local smoothness assumption $f_X\in \sH_{x_0,r}(A,\beta)$ is not sufficient in order 
 to guarantee the existence of $\widetilde{f}_X(z)$ outside this line in the complex plane.
 \item[(b)] The premise of Theorems~\ref{th:upper-bound-G1} and~\ref{th:lower-bound-G1} stipulates behavior of 
 $\widetilde{g}$ on the line $\{z: {\rm Re}(z)=1\}$ only; in particular, $\widetilde{g}(z)$ does not vanish
 on this line.
\end{itemize}
Under (a) and (b) the only possible choice of parameter $s$ is $s=0$, and
as pointed out  
in   Remark~\ref{rem:3}(ii), the form of the corresponding estimator $\hat{f}_{s,h}(x_0)$ coincides with that of 
the  deconvolution estimator 
in the additive model based on the log--transformed data. 
\par
As
discussed in Remark~\ref{rem:3}(iii), the facts 
(a) and (b) are  essential for the proof of the lower bound of Theorem~\ref{th:lower-bound-G1}, which 
is achieved on a  least favorable two--point testing problem for 
alternatives  $f_X^{(0)}$ and $f_X^{(1)}$ 
satisfying
\[
 \int_0^\infty f_X^{(i)}(x) x^{2\alpha}\rd x =\infty,\;\;\;i=0,1,\;\;\forall \alpha\ne 0.
\]
It turns out, however, that if $\widetilde{f}_X(z)$ is analytic in a strip around
$\{z: {\rm Re}(z)=1\}$ then  the upper bound of Theorem~\ref{th:upper-bound-G1}
can be improved in terms of dependence on $x_0$.  As we demonstrate below, this improvement is achieved by the choice 
of parameter $s$.

\par 
Let $\alpha>0$, $M>0$, and consider the functional class
\[
 \sF_{\alpha,M}(A,\beta) := \sH_{x_0,r}(A,\beta)\;\cap \;
 \bigg\{f: \int_0^\infty x^{2\alpha} f(x)\rd x \leq M\bigg\}.
\]
Note that for $f_X\in \sF_{\alpha, M}(A,\beta)$ it holds that 
\[
\Omega_{f_X}\supset 
\{z\in \bC: 0\leq {\rm Re}(z)\leq 2\alpha+1\}.
\]
The following statement holds.
\begin{theorem}\label{th:upper-bound-2}
For arbitrarily small $\epsilon>0,$ let 
\begin{equation}\label{eq:s-*}
s_*:=\max\big\{-\alpha, \tfrac{1}{2}(1-b)+\epsilon\big\}.
\end{equation}
Suppose that
Assumption~[G1] holds  with $\sigma=1-s_*$ and  $\gamma>1$. 
Let
$\hat{f}_{s_*, h_*}(x_0)$ be the estimator associated with kernel $K$ as in 
Theorem~\ref{th:upper-bound-G1} and  
\[
 s=s_*, 
 \;\;
 h=h_*:= C_5 \big[M^{-1}A^{2} x_0^{-2s_*+2}(x_0^\beta+1)^{2}n\big]^{-\frac{1}{2\beta+2\gamma+1}}.
\]
If $n$ is large enough so that $h_*<\min\{\ln r,1\},$
then
\begin{equation}\label{eq:R-2}
 \cR_n\big[\hat{f}_{s_*,h_*}; \sF_{\alpha,M}(A,\beta)\big] 
 \leq C_6
 [A(x_0^\beta+1)]^{\frac{2\gamma+1}{2\beta +2\gamma +1}}  
 \big(M x_0^{2s_*-2} n^{-1}\big)^{\frac{\beta}{2\beta+2\gamma+1}},
\end{equation}
where $C_6$ depends on $\beta$  only.
\end{theorem}
\begin{remark}\mbox{}
\begin{itemize}
\item[{\rm (i)}] 
If $\gamma\leq 1$ then the result of 
Theorem~\ref{th:upper-bound-2} holds for a slightly smaller
set of functions than $\sH_{x_0,r}(A,\beta)$, as discussed in Remark~\ref{rem:1}(i).
\item[{\rm (ii)}]
For $x_0$ separated away from zero by a constant, the upper bound (\ref{eq:R-2}) takes the form 
 \begin{equation}\label{eq:x0large1}
  \cR_n\big[\hat{f}_{s_*,h_*}; \sF_{\alpha,M}(A,\beta)\big] 
  \leq C_8
  A^{\frac{2\gamma+1}{2\beta +2\gamma +1}}  
 \big(M x_0^{2\gamma-1+2s_*} n^{-1}\big)^{\frac{\beta}{2\beta+2\gamma+1}}.
 \end{equation}
Because $s_*\leq 0,$ this bound is better than (\ref{eq:x0-large}) in terms of its dependence on $x_0,$ provided \(x_0>1.\)  
 For instance, let $\eta$ be uniformly distributed random variable on $[0,1]$; then $\gamma=1$, $a=0$ 
 and $b=\infty$.
 If $f_X$ has  bounded second moment, i.e., $f_X\in \sF_{1,M}(A,\beta)$, and the condition 
 in (\ref{eq:gamma<1}) holds,
 then 
 in view of (\ref{eq:s-*})
 the best choice of $s$ is $s=s_*=-1$, and the right hand
 side of (\ref{eq:x0large1}) is proportional to $x_0^{-\beta/(2\beta+3)}$. Thus,
  the  
 accuracy improves for large  $x_0$. This fact is in contrast to the result of Theorem~\ref{th:upper-bound-G1}
 stated for the functional class $\sH_{x_0,r}(A,\beta)$.
\end{itemize}
 \end{remark}
\subsubsection{Super--smooth error densities}
Now we turn to the convergence analysis of the risk of $\hat{f}_{s,h}(x_0)$ in 
the case of super--smooth error densities 
characterized by the following assumption.
\begin{enumerate}
 \item[\hbox{[G2]}] For some $\sigma \in (a,b),$ 
 there exist constants $c_0>0$, $\omega_0>0$, $\gamma>0$, $\nu\in \bR$, $B_2\geq B_1>0$   such that 
 \begin{align}
 & \;\min_{|\omega|\leq \omega_0} |\widetilde{g}(\sigma+i\omega)|\geq c_0>0,
 \nonumber
 \\*[2mm]
  & B_1 |\omega|^{\nu} e^{-\gamma|\omega|} \leq 
 |\widetilde{g}(\sigma + i\omega)| \leq 
 B_2 |\omega|^\nu e^{-\gamma |\omega|},\;\;
  \forall |\omega|\geq \omega_0.
  \label{eq:supsmooth}
 \end{align}
\end{enumerate}
The probability densities on $[0,\infty)$ 
with exponential tails are the prototypes of densities satisfying  Assumption~[G2].
\begin{example}[Gamma distribution] \label{eq:4}
 Let $g(x)=\mu^{\alpha}x^{\alpha-1}e^{-\mu x}/\Gamma(\alpha)$, $\alpha>0$, $\mu>0$, $x>0$;
 then 
 \[
  \widetilde{g}(z)= \mu^{-z+1}\Gamma(z+\alpha-1)/\Gamma(\alpha), 
  \quad {\rm Re}(z) >-\alpha+1.
 \]
As a result $a=-\alpha+1$, $b=\infty$. Furthermore, 
it is well known \cite[Corollary~1.4.4]{andrews1999special} that 
for any $\sigma\geq-2,$ there exist positive
constants $C$ and $C^\prime$ such that uniformly for $\left\vert
\omega\right\vert \geq2,$
\begin{align}
\label{gamma_asymp}
C|\omega|^{\sigma-1/2}e^{-|\omega|\pi/2}\leq\left\vert
\Gamma(\sigma+i\omega)\right\vert \leq C^\prime|\omega|^{\sigma
-1/2}e^{-|\omega|\pi/2}.
\end{align}
Thus, (\ref{eq:supsmooth}) is verified for large enough $\omega_0$ with some 
\(c_0=c_0(\omega_0)>0,\) $\nu=\sigma+\alpha-3/2$ and $\gamma=\pi/2$.
\end{example}
\begin{example}[Half--normal distribution]
\label{ex:5}
Let $g(x)=\sqrt{2/\pi} (1/\upsilon) \exp\{-x^2/(2\upsilon^2)\}$ with \(v>0.\) 
As can be easily seen, \(g(x)\) is a probability density on  $\mathbb{R}_{+}$ and it holds 
\[
 \widetilde{g}(z)= \pi^{-1/2} (\sqrt{2} \upsilon)^{z-1} \Gamma(z/2).
\]
In view of (\ref{gamma_asymp}), Assumption~[G2]  holds for large enough 
$\omega_0$ with $\nu=(\sigma-1)/2$
and $\gamma=\pi/4$.
\end{example}
\paragraph{Estimator and bounds on the risk} 
Now we analyze the accuracy of $\hat{f}_{s,h}(x_0)$ under Assumption~[G2]. In this case 
the kernel $K$ is to be constructed in a different way.
Specifically, let $\lambda\geq 2$ be a fixed natural number, and 
let $w$ be a function defined via its Fourier transform,
\begin{eqnarray}
\label{eq:hatw}
\widehat{w}(\omega)=\exp\{-|\omega|^{2\lambda}/2\lambda\}.
\end{eqnarray}
Note that $\int_{-\infty}^\infty w(x)\rd x=1$. 
For a positive integer number $m$ let
\begin{equation}\label{eq:K-supersmooth}
 K(t)=\sum_{j=1}^{m+1} \tbinom{m+1}{j} (-1)^{j+1}\tfrac{1}{j} w\big(\tfrac{t}{j}\big).
\end{equation}
It is well-known that (\ref{eq:K-supersmooth}) 
defines  kernel $K$
satisfying condition (\ref{eq:kernel-K}) (see, e.g., \cite{kerkyacharian2001nonlinear}).
Although functions $w$ and $K$ depend on the parameter $\lambda$, for the sake of brevity 
we shall not indicate this in
our notation.
For $h>0,$ let $K_{h}(x,y)$ and $L_{s,h}(x,y)$ be defined by
(\ref{eq:K-kernel}) and (\ref{eq:L-1}), respectively.
Consider the corresponding estimator 
\[
 \hat{f}_{s,h}(x_0)=\frac{1}{n}\sum_{j=1}^n L_{s,h}(x_0,Y_j).
\]
\begin{theorem}\label{th:upper-supersmooth}
Suppose that Assumption~[G2] holds with $\sigma=1$.
Let $x_0>0$, and let 
$\hat{f}_{h_*}(x_0)=\hat{f}_{0,h_*}(x_0)$ 
be the estimator 
associated with kernel $K$ given in \eqref{eq:hatw} and \eqref{eq:K-supersmooth} with parameters
\[
m\geq \lfloor \beta\rfloor +1,\;\;  
h_*= C_1\gamma \Big[\ln (A^2x_0^{2\beta+2}n)\Big]^{-1+\frac{1}{2\lambda}}.
\]
Then 
\begin{equation}\label{eq:upper-bound-G2}
\limsup_{n\to\infty} \Big\{\varphi_n^{-1} \cR_n[\hat{f}_{h_*}; \sH_{x_0, r}(A, \beta)]\Big\} \leq C_2,
\end{equation}
where $\varphi_n=A \gamma^\beta (\ln n)^{-\beta(1-\frac{1}{2\lambda})} x_0^{\beta}$, and 
$C_2=C_2(\beta, \lambda)$ depends on $\lambda$ and
$\beta$.
\end{theorem}
 \begin{remark}
  Theorem~\ref{th:upper-supersmooth} shows that for any fixed $\lambda\geq 2,$ the 
  maximal risk of $\hat{f}_{h_*}$ converges to zero at  the rate
   $O\left((\ln n)^{-\beta(1- (1/2\lambda))}\right)$ as $n\to \infty$. 
   It may seem advantageous  to let $\lambda\to \infty$ 
  as $n\to\infty$. However, 
  the constant $C_2(\beta, \lambda)$ on the right hand side of (\ref{eq:upper-bound-G2})
  explodes as $\lambda\to\infty$.
 \end{remark}
\par 
A simple modification of the proof of Theorem~\ref{th:lower-bound-G1} shows that under 
Assumption~[G2] and under suitable condition on the derivative  
$\widetilde{g}^\prime(1+i\omega)$
(similar to Assumption~[G1$^\prime$]) one has
\[
 \liminf_{n\to\infty} \Big\{\phi_n^{-1}\cR_n^*[\sH_{x_0, r}(A,\beta)]\Big\} \geq C_3,\;\;\;
 \phi_n:=A\gamma^\beta x_0^\beta (\ln n)^{-\beta},
\]
where $C_3$ depends on $\beta$ only. 
Thus the estimator $\hat{f}_{h_*}$ can be regarded as  nearly rate--optimal.
%
It is worth noting that the result of Theorem~\ref{th:upper-supersmooth} remains valid for the 
class $\sF_{\alpha, M}(A,\beta)$, and
the choice of the parameter $s\ne 0$ does not lead to improvements in
the rate of convergence 
in terms of its dependence on $x_0$.
\section{Estimation at zero}\label{sec:zero}
Now we turn to the problem 
of estimating $f_X(0)$ in the model (\ref{eq:observations}). 
The following modification of the definition of $\sH_{x_0,r}(A,\beta)$ 
will be considered.
\begin{definition} 
 Let $\beta>0$, $A>0$ and $r>0$.
We say that $f\in \sH_{r} (A,\beta),$  if 
$f$ is $\ell =\lfloor \beta\rfloor :=\max\{k\in \bN_0: k<\beta\}$ times continuously differentiable on \((0,r]\) 
and $\max_{k=1,\ldots, \ell} |f^{(k)}(x)|\leq A$,
\[
 \big|f^{(\ell)}(x)-f^{(\ell)}(x^\prime)\big|\leq 
 A|x-x^\prime|^{\beta -\ell},\;
 \forall x, x^\prime \in (0, r].
\]
We define also 
\begin{eqnarray}
\label{eq:def-bar-H}
 \bar{\sH_{r}} (A,\beta,M) := \sH_{r} (A,\beta) \cap \Big\{f: \sup_{t\in \mathbb{R}_{+}}|f(t)|\leq M\Big\}.
\end{eqnarray}
\end{definition} 
First we note that if $I_g:=\int_0^\infty [g(x)/x]\rd x <\infty$, i.e., if $\{z: {\rm Re}(z)=0\}\subseteq \Omega_g,$ then $f_Y$
is finite at the origin, and 
in view of~(\ref{eq:f_Y})
$f_Y(0)= f_X(0)I_g$.
In this case a natural estimator of $f_X$ can be defined as
$\hat{f}_X(0)=\hat{f}_Y(0)/I_g$,  
where 
$\hat{f}_Y(0)$ is a suitable estimator of $f_Y(0)$, say, a kernel-type estimator with bandwidth $h$, 
from direct observations $Y_1,\ldots, Y_n$.
As a result, under the choice   \(h\asymp n^{-1/(2\beta+1)}\) 
(see e.g. Theorem 1.1 in \cite{tsybakov2009introduction}), we get
\begin{eqnarray*}
\cR_n\big[\hat{f}_{h}; \sH_{r}(A,\beta)\big] \leq O( n^{-\beta/(2\beta+1)}). 
\end{eqnarray*}
It is also clear that this rate is minimax over the class 
\(\sH_{r}(A,\beta).\) 
Note, however, that the condition $\{z: {\rm Re}(z)=0\}\subseteq \Omega_g$ 
is too restrictive 
and does not hold in many situations of interest. 
For instance, it does not hold for the uniform
distribution on $[0,1]$. Thus, in  the case when 
$\{z: {\rm Re}(z)=0\}$ is not a subset of  $\Omega_g$,
we need to propose an alternative method of estimating \(f_X(0).\)
\subsection{Kernel construction and estimator}
In order to construct an estimator of \(f\) at zero, 
we use the following kernel. 
For a fixed real number $s\geq 0,$ consider the function
\begin{equation}\label{eq:psi-sigma}
 \psi_s(x)=\tfrac{1}{\sqrt{2\pi}} e^{-\frac{1}{2}(1-s)^2} x^{-s} 
 \exp\{-\tfrac{1}{2}[\ln x]^2\},\;\;
 x\geq 0.
\end{equation}
It is easily checked that $\int_0^\infty \psi_s(x)\rd x=1$ and 
$\widetilde{\psi}_s(s+i\omega)=\frac{1}{\sqrt{2\pi}} e^{-\frac{1}{2}(1-s)^2} e^{-\frac{1}{2}|\omega|^2}$. 
Fix   positive integer number $m$, and 
define the kernel
\begin{equation}\label{eq:K-sigma}
 K_s (x)= \sum_{j=1}^{m+1} \tbinom{m+1}{j} (-1)^{j+1} \tfrac{1}{j}\psi_s \big(\tfrac{x}{j}\big),
 \;\;\;x\geq 0.
\end{equation}
By construction, $K_s$ satisfies condition (\ref{eq:kernel-K}). Another attractive property
of the kernel $K$ 
 is that the Mellin transform $\widetilde{K}_s(z)$ decreases
at the rate $e^{-\frac{1}{2}|\omega|^2}$ as $|\omega|\to \infty$ along the line $\{z:{\rm Re}(z)=s\}$  
[see the proof of Theorem~\ref{thm:upper_zero}].
\par
Having defined the function $K_s$,  let us consider its scaled version, $K_{s, h}(x):=(1/h) K_s(x/h)$
for $h>0$, and note that 
\[
 \widetilde{K}_{s,h}(z)=\int_0^\infty t^{z-1} K_{s,h}(t)\rd t= h^{z-1} \widetilde{K}_s(z).
\] 
The kernel $L_{s,h}(y)$ corresponding  to $K_{s,h}(x)$  
is given by  
\begin{eqnarray}
 L_{s,h}(y) &:=& \frac{1}{2\pi i} \int_{s-i\infty}^{s+i\infty} 
 \frac{\widetilde{K}_{s,h}(z)}{\widetilde{g}(1-z)} y^{-z} \rd z
\nonumber
 \\
& =& \frac{1}{2\pi h^{1-s} y^s}\int_{-\infty}^\infty \Big(\frac{h}{y}\Big)^{i\omega} 
\frac{\widetilde{K}_s(s+i\omega)}{\widetilde{g}(1-s-i\omega)}
\rd \omega, 
\label{eq:Lsh}
 \end{eqnarray}
 provided that  the expression on the right hand side
is well defined. 
\par 
Consider now the following estimator 
\begin{equation}\label{eq:Lh-est-2}
 \hat{f}_{s,h}(0) = \frac{1}{n} \sum_{i=1}^n L_{s,h}(Y_i).
\end{equation}
The tuning parameters $s$ and $h$ 
will be specified below in   
Theorem~\ref{thm:upper_zero}.

\subsection{Bounds on the risk}

First we establish an upper bound on the maximal risk of the estimator $\hat{f}_{s,h}(0)$. 
It is done under the following assumptions on the error density $g$.
\begin{itemize}
 \item [\hbox{[G3]}] 
 For some $p\in [0, 1)$, $q\geq 0$ and $\delta\in (0, 1)$
 \begin{equation}\label{eq:g-near-zero}
 c_0 x^{-p} [\ln (1/x)]^{q} \leq  g(x) \leq C_0 x^{-p} [\ln (1/x)]^{q},\quad  x\in (0, \delta).
 \end{equation}
\end{itemize}
\par 
Assumption [G3] prescribes behavior of the density 
 $g$ in a vicinity of the origin. If $p<0$ then the integral $\int_0^\infty [g(x)/x]\rd x$ is finite, 
 and, as discussed above, the problem reduces to the density estimation from
 direct observations. Moreover, since $g$ is a probability density, it must hold $p<1$.
That is why in [G3] we restrict our attention  to the case $p\in [0,1)$. 
Note also that [G3] implies that $\widetilde{g}$ is well defined in the strip $\{z: p<{\rm Re}(z)\leq 1\}$, i.e.,
$\Omega_g \supseteq \{z: p<{\rm Re}(z)\leq 1\}$.
\par
In addition to Assumption~[G3], we impose some mild conditions on $g$  that guarantee existence 
of the estimator $\hat{f}_{s,h}(0)$ under the following specific choice of the parameter $s$, 
\begin{equation}\label{eq:s*}
 s_*:=\tfrac{1}{2}(1-p);
\end{equation}
here $p$ is the parameter appearing in Assumption~[G3].
\begin{itemize}
 \item[\hbox{[G4]}] Suppose that $|\widetilde{g}(1-s_*+i\omega)|>0$ for all $\omega\in \bR$,
 and 
 \begin{equation}\label{eq:G4-1}
 \int_{-\infty}^\infty \frac{e^{-\omega^2/2}}{|\widetilde{g}(1-s_*+i\omega)|}\rd \omega\;\vee\;
 \int_{-\infty}^\infty \frac{e^{-\omega^2}}{|\widetilde{g}(1-s_*+i\omega)|^2}\rd \omega
 \leq C_1 <\infty.
 \end{equation}
In addition,
\begin{equation}\label{eq:G-4-2}
 \int_{-\infty}^\infty \bigg|\frac{\rd^l}{\rd \omega^l} 
 \bigg(\frac{e^{-\omega^2/2}}{\widetilde{g}(1-s_*+i\omega)}\bigg) \bigg|^2 \rd \omega \leq C_2 <\infty,
\end{equation}
where $l:=\lceil (q+1)/2\rceil$, and $q$ appears in (\ref{eq:g-near-zero}).
\end{itemize}
\par 
The conditions of Assumption~[G4] are rather mild.
First we note that under Assumption~[G3] the line 
$\{z: {\rm Re}(z)=1-s_*=\frac{1}{2}(1+p)\}$ belongs  to the convergence region of $\widetilde{g}$. 
The first condition in~[G4]  bounds from below the rate of decay of $\widetilde{g}$
along this line.
 It ensures that under the choice $s=s_*$ the integrand 
 in (\ref{eq:Lsh}) is absolutely integrable and square integrable; 
 thus the estimator $\hat{f}_{s_*,h}(0)$
in (\ref{eq:Lh-est-2}) is well defined [see the proof of Theorem~\ref{thm:upper_zero} for details].
 The second condition of [G4] is stated for the derivatives of the integrand in (\ref{eq:Lsh})
 and is used to bound the variance of $\hat{f}_{s_*,h}(0)$. Note that (\ref{eq:G4-1}) holds both for the smooth and 
 super--smooth error densities. 
 \par
We are now in a position to state an upper bound on the risk of the estimator $\hat{f}_{s_*,h}(0)$
under a suitable choice of the bandwidth $h$.
\begin{theorem}
\label{thm:upper_zero}
 Fix some positive real numbers \(A,\) \(\beta,\)  \(M\) and 
 consider the class of functions \(\bar{\sH}_{r}(A,\beta,M)\) defined in \eqref{eq:def-bar-H}. Let  Assumptions~[G3] and [G4] hold, and let $\hat{f}_*(0)=\hat{f}_{s_*,h_*}(0)$ denote  
 the estimator \eqref{eq:Lh-est-2} associated with parameters
 $m\geq \lfloor \beta\rfloor+1$,  $s=s_*$ given by (\ref{eq:s*}) and 
 \begin{equation}\label{eq:kappa}
 h=h_*:= \big[MA^{-2}(\ln n)^{q+\kappa}n^{-1}\big]^{\frac{1}{2\beta+1+p}},
  \;\;\;
  \kappa:=\left\{\begin{array}{ll}
                  0, & p\in (0,1),\\
                  1, & p=0.
                 \end{array}
\right.
 \end{equation}
Then for $n$ large  enough such that $h_*<\min\{r, 1\}$ one has
\[
 \cR_n[\hat{f}_*; \bar{\sH}_{r}(A,\beta,M) ] \leq C_3 
 A^{\frac{1+p}{2\beta+1+p}}\big[M(\ln n)^{q+\kappa}n^{-1} \big]^{\frac{\beta}{2\beta+1+p}},
\]
where $C_3$ may depend on $\beta$ only.
\end{theorem}

\begin{remark}
 \mbox{}
 \begin{itemize}
 \item[{\rm (i)}] Note that the upper bound of Theorem~\ref{thm:upper_zero} holds both for smooth and super--smooth error densities,
 provided that the mild conditions of Assumption [G4] are fulfilled.
 This is in  contrast to the results on estimating 
 density $f_X$ at a point separated away from zero.
 \item[{\rm (ii)}] 
 It is instructive to consider 
 particular cases corresponding to different error densities. For instance, if 
 $g$  is the uniform density  on $[0,1]$, or an exponential density 
 then
 $p=0$, $q=0$ and $\kappa=1$. So in these cases 
 the upper bound is of the order $(\ln n/n)^{\beta/(2\beta+1)}$ which is only
 by a logarithmic factor worse
 than the standard nonparametric rate.
 \end{itemize}
\end{remark}
\par 
Our next result is the lower bound on the minimax risk. 
To that end, we introduce the following condition~on~$g$.
\begin{itemize}
\item[\hbox{[G5]}] 
Suppose that 
$\{z\in \bC: 1\leq {\rm Re}(z)\leq 1+\epsilon\} \subset \Omega_g$ for some $\epsilon>0$,
and 
 \begin{equation}\label{eq:moment-g}
|\widetilde{g}(1+\epsilon +i\omega)|
\leq C_4<\infty,\;\;\;\forall \omega.
\end{equation}
\end{itemize}
\par 
Assumption~[G5] is rather mild; it holds if $\int_0^\infty x^\epsilon g(x)\rd x \leq C_4$ for some $\epsilon>0$.
Note also that~[G5] together with  [G3] imply that $\widetilde{g}$ is analytic in 
the strip $\{z: p<{\rm Re}(z)\leq 1+\epsilon\}$.
\begin{theorem}\label{th:lower-bound-zero}
Let Assumptions~[G3] and [G5] hold, then  for the functional class $\bar{\sH}_r(A, \beta, M)$ with $M\geq 1$ one has
\[
 \liminf_{n\to\infty} \;\Big\{\,\phi_n^{-1} \cR_n^*[\bar{\sH}_{r}(A,\beta,M)]\,\Big\} \geq C_5, 
 \]
 where
 \[
 \phi_n:= A^{\frac{p+1}{2\beta+1+p}}
\big[M^{1-p} (\ln n)^{q+\kappa} n^{-1} \big]^{\frac{\beta}{2\beta+1+p}},
\]
and  
$C_5$ depends on $\beta$ only.
\end{theorem}
The lower bound on the minimax risk of Theorem~\ref{th:lower-bound-zero} matches
the bound of 
Theorem~\ref{thm:upper_zero} up to a minor discrepancy in terms of  dependence on $M$. 
Note, however,
that in the practically important case of $p=0$ the  bounds coincide. Thus the estimator 
$\hat{f}_{*}(0)$ is rate--optimal on the class $\bar{\sH}_r(A,\beta, M)$.

\section{Numerical experiments}
\label{sec:sim}
In this section we demonstrate that in 
many cases of interest the developed estimators are given by analytic formulas 
and can be easily implemented. We also illustrate numerically theoretical results on performance 
of the estimators.

\subsection{Estimation outside zero}\label{sec:sim-1}
First we study numerically the accuracy of the estimator \eqref{eq:estimator} for 
points separated away from zero.
Assume that errors  $(\eta_i)$  are beta--distributed with the density
\begin{eqnarray}
\label{beta_dens}
g(x)=\nu x^{\nu-1},\quad  0\leq x\leq 1,\;\;\nu>0,
\end{eqnarray}
then
\begin{eqnarray}\label{eq:beta_dens1}
\widetilde{g}(z)=\nu\int_{0}^{1}x{}^{\nu-1}x^{z-1}\,dx=\nu/(\nu+z-1).
\end{eqnarray}
Furthermore, consider the case of  exponentially distributed \(X\), 
that is, \(f_X(x)=e^{-x}\) for \(x>0.\) Let $w(x)=e^{-x^{2}/2}/\sqrt{2\pi}$, and for a fixed  natural number $m$ 
let
\begin{equation}\label{eq:K-Nikolski}
K(t)=\sum_{j=1}^{m+1}\tbinom{m+1}{j}(-1)^{j+1}\tfrac{1}{j}w\big(\tfrac{t}{j}\big).
\end{equation}
The bilateral Laplace transform of $K$ is defined for any $z\in\mathbb{C}$ and given by
\begin{multline*}
\widecheck{K}(z) 
=
\sum_{j=1}^{m+1}\tbinom{m+1}{j} 
(-1)^{j+1}\frac{1}{\sqrt{2\pi}j}\int_{-\infty}^{\infty}e^{-t^{2}/(2j^2)-tz}\,\rd t
\\
=\sum_{j=1}^{m+1}\tbinom{m+1}{j}(-1)^{j+1}e^{j^2z^{2}/2}.
\end{multline*}
Let us now compute the kernel $L_{s,h}(x,y)$,
\[
L_{s,h}(x,y):=\frac{1}{2\pi ix}\int_{s-i\infty}^{s+i\infty}
\left(\frac{x}{y}\right)^{z}\,\frac{\widecheck{K}(zh)}{\widetilde{g}(1-z)}\,\rd z.
\]
%
Using (\ref{eq:beta_dens1}), we obtain
%
\begin{multline*}
\frac{1}{2\pi ix}\int_{s-i\infty}^{s+i\infty}
\left(\frac{x}{y}\right)^{z}\,\frac{e^{j^2h^{2}z^{2}/2}}{\widetilde{g}(1-z)}\,\rd z 
\\
  =\frac{1}{2\pi\nu x^{1-s}y^{s}}\int_{-\infty}^{\infty}e^{iu\ln(x/y)}\,(\nu-s-iu)e^{j^2h^{2}(s+iu)^{2}/2}\,\rd u\\
 =
 \frac{1}{\sqrt{2\pi}x^{1-s}y^{s}}
 \exp\bigg\{\frac{j^2s^{2}h^{2}}{2} -\frac{1}{2j^2h^2}[j^2sh^{2}+\ln(x/y)]^{2}\bigg\}
 \\
 \times
 \bigg\{\frac{\nu-s}{(j^2h^2)^{1/2}}+\frac{j^2sh^{2}+\ln(x/y)}{(j^2h^{2})^{3/2}}\bigg\}.
\end{multline*}
Thus
\begin{align*}
L_{s,h}(x,y) 
 & =\frac{1}{\sqrt{2\pi}}\sum_{j=1}^{m+1}\tbinom{m+1}{j} 
 (-1)^{j+1}
 \exp\Big\{-\frac{\ln^{2}(x/y)}{2j^2h^{2}}
 \Big\}\frac{1}{xjh}\Big[\nu+\frac{\ln(x/y)}{j^2h^{2}}\Big].
\end{align*}
Note that the kernel does not depend on \(s\) and this corresponds to the fact that the function    \(\widecheck{K}(zh)/\widetilde{g}(1-z)\) is holomorphic. 
\par
\begin{figure}
\begin{center}
\includegraphics[width=0.45\linewidth ]{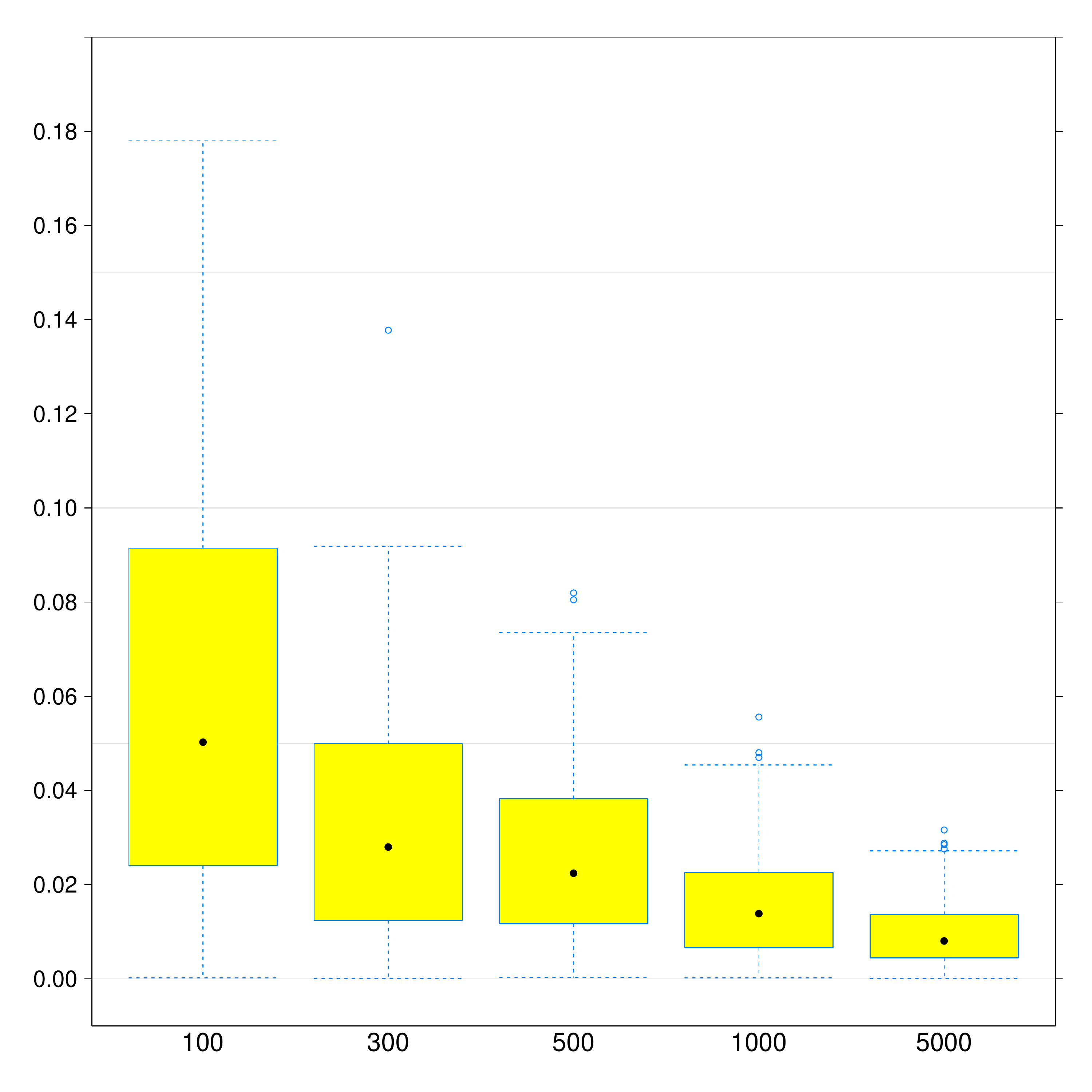}
\includegraphics[width=0.45\linewidth ]{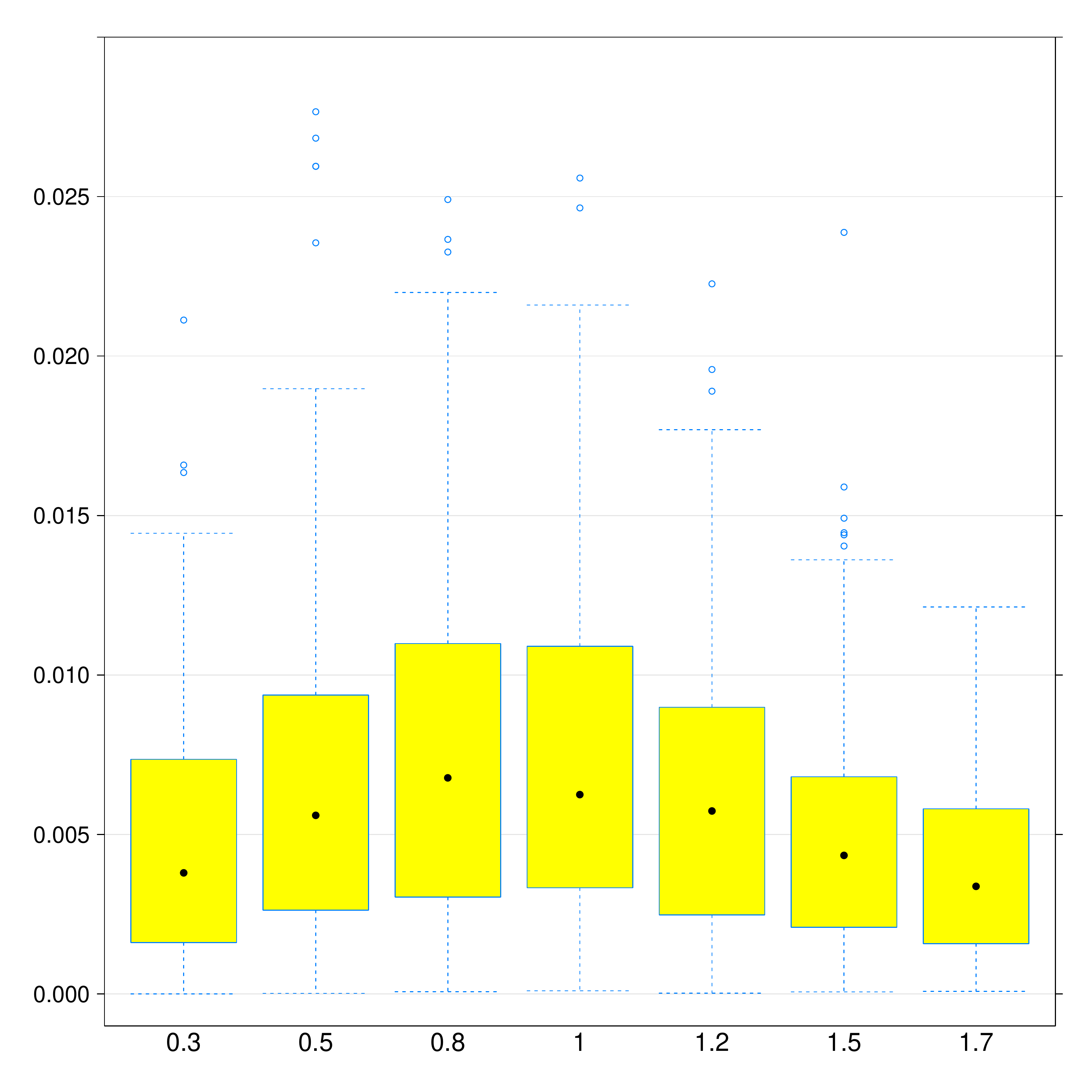}
\caption{\label{fig2} Left: boxplots of the distance  \(|\hat{f}_{h_\star}(1)-f_X(1)|\), where the estimate   \(\hat{f}_{h_\star}(1)\) is
based on \(n\in \{100, 300, 500, 1000,5000\}\) observations of the r.v. \(Y\) under uniformly distributed errors. Right: boxplots of the distance  \(|\hat{f}_{h_\star}(x)-f_X(x)|\) for \(x\in \{0.3,0.5,0.8,1.0,1.2,1.5,1.7\},\) where the estimate   \(\hat{f}_{h_\star}(x)\) is
based on \(n=500\) observations of the r.v. \(Y\) under uniformly distributed errors. The bandwidth \(h_\star\) is precomputed using \(300\) independent runs.}
\end{center}
\end{figure}

In Figure~\ref{fig2}  we present box plots of the quantity \(|\hat{f}_{h_\star}(x)-f_X(x)|\) 
for different sample sizes \(n\) and 
different points \(x>0\)  over \(200\) simulation runs, 
where in each run we construct the estimate \(\hat{f}_{h_\star}(x)\)
associated with  the above kernel $L_{s,h}$ and a precomputed bandwidth \(h_\star\). The 
 latter is found by minimizing 
 \(\bE_N[|\hat{f}_{h}(x)-f_X(x)|^2]\) over \(h\) 
 with the empirical expectation \(\bE_N\) computed using \(N=300\) independent simulation runs. 
 The left graph in Figure~\ref{fig2} demonstrates convergence of the estimation error for $x_0=1$  as the sample sample grows,
 while the right graph shows dependence of the error for a given sample size $n=500$ on $x_0$.  
 As can be seen  the error decreases as $x_0$ grows, which is 
 in accordance with the results of Theorem~\ref{th:upper-bound-2}.

\subsection{Estimation at zero}
Now we illustrate behavior of the developed estimator for the case  $x_0=0$.
We consider again beta--distributed errors as in  (\ref{beta_dens}) and (\ref{eq:beta_dens1}).
%
Let  \(w(x)=e^{-x}\), and let $K$ be given by (\ref{eq:K-Nikolski}).
Using the fact that \(\widetilde w(z)=\Gamma(z)\), we have for any 
$s> \max(0, \nu)$
\begin{eqnarray*}
 \frac{1}{2\pi}\int_{-\infty}^\infty e^{-i\omega y} \frac{\widetilde{w}(s+i\omega)}{\widetilde{g}(1-s-i\omega)}\rd \omega 
 = 
 \frac{1}{2\pi} \int_{-\infty}^\infty e^{-i\omega y} \Gamma(s+i\omega)\Big(1-\frac{s+i\omega}{\nu}\Big)\rd \omega
\\
= \frac{1}{2\pi} \int_{-\infty}^\infty e^{-i\omega y} \Gamma(s+i\omega)\rd \omega - 
\frac{1}{2\pi\nu} \int_{-\infty}^\infty e^{-i\omega y} \Gamma(1+s+i\omega)\rd \omega. 
 \end{eqnarray*}
The well-known identity 
\[
\frac{1}{2\pi}\int_{-\infty}^{\infty}e^{-i\omega y}\Gamma(s+i\omega)\,\rd\omega
=e^{sy}\exp\left(-e^{y}\right), \quad y\in \mathbb{R}
\]
leads to  
\begin{align*}
\frac{1}{2\pi}\int_{-\infty}^\infty e^{-i\omega y}
\left(1-\frac{s+i\omega}{\nu}\right)\Gamma(s+i\omega)\,\rd\omega &  
=e^{sy}\exp\left(-e^{y}\right)\left(1-\frac{e^{y}}{\nu}\right).
\end{align*}
Then using  (\ref{eq:Lsh}) and a straightforward algebra, we obtain 
\[
L_{s,h}(y)=\sum_{j=1}^{m+1}\tbinom{m+1}{j}(-1)^{j+1}\frac{1}{jh}
\exp\Big\{-\frac{y}{jh}\Big\}\Big(1-\frac{y}{jh\nu}\Big).
\]
The corresponding estimator is
$\hat{f}_{h}(0) := \frac{1}{n} \sum_{i=1}^n L_{s,h}(Y_i)$.
\par
\begin{figure}
\begin{center}
\includegraphics[width=0.45\linewidth ]{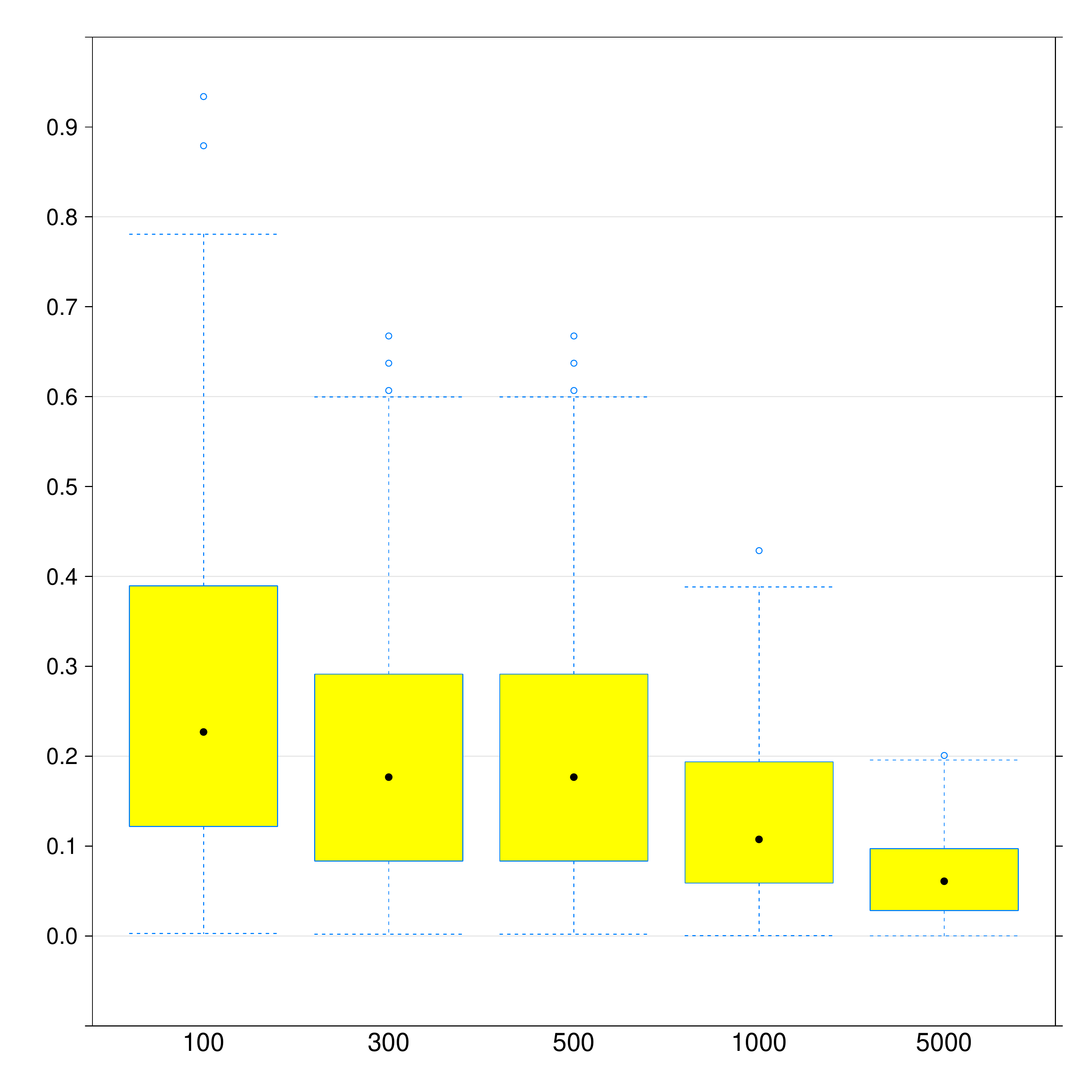}
\includegraphics[width=0.45\linewidth ]{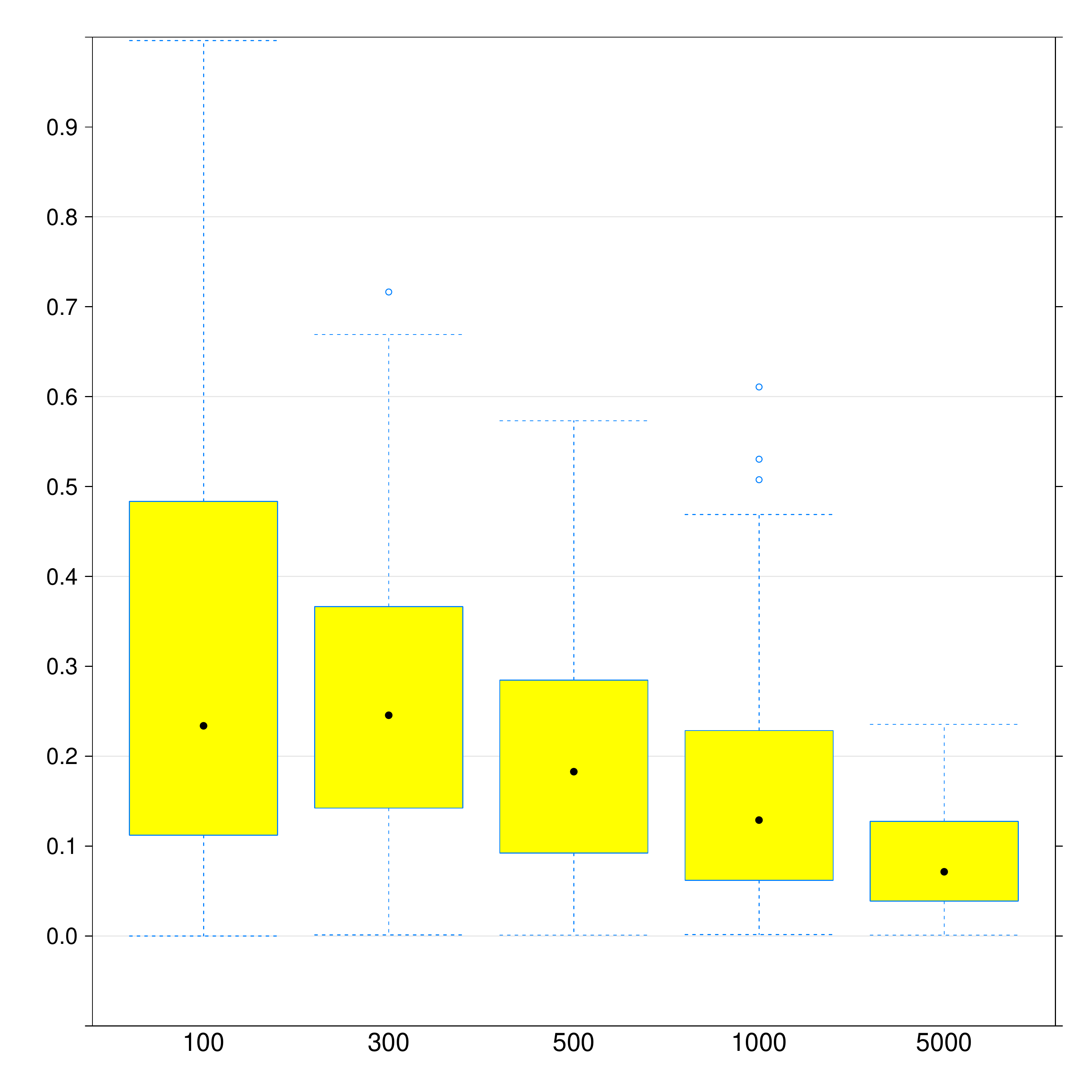}
\caption{\label{fig1} Boxplots of the distance  \(|\hat{f}_{h}(0)-f_X(0)|\), where the estimate   \(\hat{f}_{h}(0)\) is
based on \(n\in \{100, 300, 500, 1000,5000\}\) observations of the r.v. \(Y\) 
under beta-distributed errors with density \eqref{beta_dens} with parameters \(\nu=1\) (left) and \(\nu=1/2\) (right).}
\end{center}
\end{figure}

In our simulation study we take \(f_X(x)=2\exp(-2x)\) so that \(f_X(0)=2\)
and the distribution of \(\eta\) as in \eqref{beta_dens} with \(\nu\in \{1,\frac{1}{2}\}\).
 In Figure~\ref{fig1} we present box plots of the quantity \(|\hat{f}_{h}(0)-f_X(0)|\) over \(200\) simulation runs, where in each run we construct the estimate \(\hat{f}_{h_\star}(0)\)
 using a precomputed bandwidth \(h_\star\). 
 The latter is found by minimizing \(\bE_N[|\hat{f}_{h_\star}(0)-f_X(0)|^2]\) 
 over \(h\) with empirical expectation \(\bE_N\) computed using \(N=300\) independent simulation runs. 
As expected, in the case  $\nu=1$ the estimator is more accurate than in the case $\nu=1/2$. 
\section{Proofs of main results}
\label{sec:proofs}
In the proofs below $c_0,  c_1, c_2,\ldots$ denote positive constants
depending on the parameters 
appearing in Assumptions~[G1]--[G5] and  on $\beta$ only unless specified otherwise. 
\subsection{Proof of Theorem~\ref{th:upper-bound-G1}}
Note that under Assumption~[G1] condition (\ref{eq:kernel-K-2}) with $q>\gamma+1$ guarantees that the estimator 
$\hat{f}_h(x_0)=\hat{f}_{0,h}(x_0)$ is well--defined. Indeed, under this condition 
$\widehat{K}(\cdot h)/\widetilde{g}(1-i\cdot)\in \bL_1(\bR)\cap \bL_2(\bR)$. 
\par\medskip 
1$^0$. The next statement establishes an upper bound on the bias of $\hat{f}_{s,h}(x_0)$.
\begin{lemma}\label{lem:bias} 
Let $K_h(\cdot, \cdot)$
be given  by (\ref{eq:K-kernel}), where  $K$ satisfies (\ref{eq:kernel-K}) with 
$m\geq \lfloor \beta\rfloor +1$; then for any $x > 0$ and $h\in (0,\ln r)$
 \[
  \sup_{f\in \sH_{x,r} (A,\beta)} \Big|\int_{-\infty}^\infty K_h(x,y) f(y)\rd y - f(x)\Big| \leq 
  c_0 A\|K\|_1\Big[ h^\beta |x|^\beta + h^{\ell+1}\sum_{k=0}^\ell |x|^k\Big],
 \]
 where $c_0$ depends on $\beta$ only, and $\|K\|_1=\int_{-1}^1 |K(x)|\rd x$.
\end{lemma}
The proof of Lemma~\ref{lem:bias} is given in Section~\ref{sec:auxiliary}.
\par\medskip 
2$^0$. Now we derive an upper bound on the variance.
  Using the Cauchy--Schwarz inequality  we obtain
\begin{multline*}
 \bE_{f_X}\big[ L_h^2(x_0, Y_j)\big] 
 \\
 =  
 \frac{1}{4\pi^2 x_0^2} \int_{-\infty}^\infty \int_{-\infty}^\infty |x_0|^{i(\omega-\mu)} 
 \widetilde{f}_Y(1-i(\omega-\mu))\frac{\widehat{K}(\omega h)}{\widetilde{g}(1-i\omega)}
 \cdot \overline{\frac{\widehat{K}(\mu h)}{\widetilde{g}(1-i\mu)}}\rd \omega \rd \mu
 \\*[2mm]
 \leq\;
 \frac{1}{4\pi^2x_0^2} 
 \int_{-\infty}^\infty  |\widetilde{f}_Y(1-i\mu)| \rd \mu 
 \int_{-\infty}^\infty
 \frac{
  |\widehat{K}(\omega h)|^2}
 {|\widetilde{g}(1-i\omega)|^2}\rd \omega.
 \end{multline*}
 If $\gamma>1$ then $\widetilde{f}_Y(1-i\mu)$ is integrable:
 \begin{multline*}
  \int_{-\infty}^\infty |\widetilde{f}_Y(1-i\mu)|\rd \mu = 
  \int_{-\infty}^\infty |\widetilde{f}_X(1-i\mu)| 
  \cdot |\widetilde{g}(1-i\mu)| \rd \mu
  \\
  \leq \int_{-\infty}^\infty    
  |\widetilde{g}(1-i\mu)| \rd \mu \leq c_1 <\infty,
 \end{multline*}
 where the upper bound in (\ref{eq:as1-2}) has been used.
 Moreover, in view of (\ref{eq:kernel-K-2}) and the lower bound in~(\ref{eq:as1-2})
 we have 
 \[
 \int_{-\infty}^\infty
 \frac{
  |\widehat{K}(\omega h)|^2}
 {|\widetilde{g}(1-i\omega)|^2}\rd \omega\;\leq\; c_2 h^{-2\gamma-1}.
 \]
 Combining these bounds we obtain 
 ${\rm var}_{f_X}\{\hat{f}_h(x_0)\} \leq c_3 x_0^{-2} h^{-2\gamma-1}n^{-1}$.
\par 
On the other hand, Lemma~\ref{lem:bias} 
and $h\leq 1$ imply that  
\[ 
\sup_{f_X\in \sH_{x_0, r}(A,\beta)} \big|\bE_{f_X} [\hat{f}_{h}(x_0)] - f_{X}(x_0)\big|\leq c_3A (x_0^\beta + 
1)h^{\beta}.
\]
Then (\ref{eq:R-1}) follows from substitution of $h_*$ in the bounds for the bias and the variance.
\epr

\subsection{Proof of Theorem~\ref{th:lower-bound-G1}}
The proof is based on the standard technique for proving lower bounds (see 
\cite[Chapter~2]{tsybakov2009introduction}).
Recall that 
for two generic functions $u$ and $w$ on $[0,\infty)$ 
we write $[w\star u] (y):=\int_0^\infty (1/x) w(x)u(y/x)\rd x$. 
\par 
0$^0$. 
Let $\psi:\bR\to\bR$ be a function such that its Fourier transform $\widehat{\psi}$ is an infinitely differentiable
function satisfying 
for some $\delta\in (0,\tfrac{1}{4})$ 
\[
 \cF[\psi; \omega]= \widehat{\psi}(\omega)=\left\{\begin{array}{ll}
1, &      \omega\in [-2+\delta, -1-\delta]\cup [1+\delta, 2-\delta],
\\
0, & \omega \in (-\infty, -2]\cup [-1,1] \cup [2,\infty).
                               \end{array}
\right.
\]
Let $x_0\geq c_0>0$ for some constant $c_0$, and define 
\begin{equation*}
 f_X^{(0)}(x):=\frac{1}{\pi x [1+\ln^2(x/x_0)]},\;\;\;x>0.
\end{equation*}
Define 
\begin{equation*}
 f_X^{(1)}(x)=f_X^{(0)}(x) + \theta \psi_h(x),\;\;\;\;
 \psi_h(x):=\frac{1}{x}\psi\bigg(\frac{\ln (x/x_0)}{h}\bigg),
\end{equation*}
where $h\in (0,1)$ and $\theta>0$ are the parameters to be specified. 
\par 
1$^0$. First we show that if $\theta$ is small enough, $\theta \leq \min\{ \tfrac{1}{2} , c_1h^{-2}\}$ 
then $f_X^{(1)}$ is a probability density on $[0,\infty)$.
Indeed,   since $\widehat{\psi}(0)=0$  
\[
 \int_0^\infty \psi_h(x)\rd x = \int_0^\infty \frac{1}{x}\psi\bigg(\frac{\ln(x/x_0)}{h}\bigg)\rd x =
 h\int_{-\infty}^\infty \psi(t)\rd t=0.
\]
Thus, $f_X^{(1)}$ integrates to one. 
Moreover, 
by construction $\psi$ is rapidly decreasing as $t\to\infty$; in particular,
$|\psi(t)| \leq \pi^{-1}\min\{1, c_1t^{-2}\}$,  $\forall t\in \bR$
with some absolute constant $c_1$. Therefore, the conditions  
$\theta\leq \frac{1}{2}$ and $c_1\theta h^2\leq \frac{1}{2}$ imply that 
$\theta |\psi(t/h)| \leq [\pi (1+t^2)]^{-1}$, which, in turn shows that $f_X^{(1)}$ is non--negative.
Therefore $f_X^{(1)}$ is the probability density.
\par\medskip 
2$^0$.
First we note that if $x_0\geq c_0>0$ for some $c_0$ large enough then 
$f_X^{(0)}\in \sH_{x_0,r}(A/2, \beta)$.
Now we show that if $\theta=c_2 A x_0^{\beta+1}h^\beta$ for some constant $c_2$ then 
 $f_X^{(1)}\in \sH_{x_0, r} (A,\beta)$.
\par
For simplicity and without loss of generality assume that $\beta$ is integer, $\beta\geq 1$. 
Then by the Fa\'a~di~Bruno formula
\begin{multline*}
 \psi_h^{(\beta)}(x) = 
 \sum_{j=0}^{\beta} \binom{\beta}{j} \frac{(-1)^j}{x^{j+1}} \frac{d^{\beta-j}}{d x^{\beta-j}}
 \psi\bigg(
 \frac{\ln (x/x_0)}{h}\bigg)
 \\
  =  \sum_{j=0}^{\beta} \binom{\beta}{j} \frac{(-1)^j}{x^{j+1}}
 \sum \frac{(\beta-j)!}{k_1!\cdots k_{\beta-j}!} \psi^{(k)}\bigg(
 \frac{\ln (x/x_0)}{h}\bigg) h^{-k} x^{-(\beta-j)} \prod_{i=1}^{\beta-j} \bigg[
 \frac{(-1)^{i+1}}{i!}\bigg]^{k_i},
\end{multline*}
where the second summation is over all partitions of $\beta-j$, and  $k:=k_1+\ldots+k_{\beta-j}$, 
$k_1+2k_2+\cdots+(\beta-j)k_{\beta-j}=\beta-j$.
It follows from this expression and the fact that $h<1$ that
\[
 |\psi_h^{(\beta)}(x)|\leq c_3 x^{-\beta-1} h^{-\beta} \max_{k=1,\ldots, \beta} 
 \bigg|\psi^{(k)}\bigg(\frac{\ln (x/x_0)}{h}\bigg)\bigg|,\;\;\;\forall x>0,
\]
where $c_3$ depends on $\beta$ only.
Since $\psi$ is an infinite differentiable rapidly decreasing function, we obtain
\[
 |\psi^{(\beta)}_h(x)| \leq c_4 x_0^{-\beta-1} h^{-\beta}, \;\;\;\;\;r^{-1}x_0\leq x\leq rx_0,
\]
where $c_4$ depends on $\beta$.
 Then setting $\theta=c_2 A x_0^{\beta+1}h^\beta$, by choice of $c_2$ we obtain 
$f_X^{(1)}\in \sH_{x_0,r}(A, \beta)$.
\par\medskip 
3$^0$. Next we bound the $\chi^2$--divergence between $f^{(1)}_Y$ and $f^{(0)}_Y$. 
We have 
\begin{eqnarray*}
 f_Y^{(0)}(y) &=& [f_X^{(0)}\star g] (y) = \frac{1}{\pi y}\int_0^\infty \frac{g(x)}{1+[\ln(y/x_0)-\ln (x)]^2}
 \rd x
 \\
 &\geq& \frac{1}{\pi y[1+2\ln^2(y/x_0)]}\int_0^\infty \frac{g(x)}{1+2\ln^2(x)}\rd x \geq \frac{c_5}{y[1+2\ln^2(y/x_0)]}.
\end{eqnarray*}
Furthermore, 
\begin{eqnarray}
 f_Y^{(1)}(y) - f_Y^{(0)}(y) &=& \theta [g\star \psi_h](y) = \theta \int_0^\infty
 \frac{1}{x} g(x) \psi_h(y/x)\rd x
 \nonumber
 \\
 &=& \frac{\theta}{2\pi y}\int_{-\infty}^\infty \widetilde{g}(1+i\omega) 
 \widetilde{\psi}_h(1+i\omega) y^{-i\omega}\rd \omega,
\label{eq:f1-f0}
 \end{eqnarray}
where in the second line we have applied the inverse Mellin transform formula.
By definition of $\psi_h$,  
\begin{multline*}
 \widetilde{\psi}_h(1+i\omega)=
 \int_0^\infty x^{i\omega}\psi_h(x)\rd x = \int_0^\infty x^{i\omega-1}\psi
 \bigg(\frac{\ln (x/x_0)}{h}\bigg)\rd x
 \\
 =  h x_0^{i\omega}\int_{-\infty}^\infty e^{ith\omega} \psi(t) \rd t = 
 h x_0^{i\omega} \widehat{\psi}(-\omega h).
\end{multline*}
Substituting this expression in (\ref{eq:f1-f0})
we obtain 
\begin{eqnarray*}
 f_Y^{(1)}(y)-f_Y^{(0)}(y)  = 
 \frac{\theta h}{2\pi y} \int_{-\infty}^\infty \widetilde{g}(1+i\omega) \widehat{\psi}(-\omega h)
 e^{-i\omega \ln (y/x_0)}\rd \omega=: \frac{\theta h}{2\pi y} \rho \big(\ln (y/x_0)). 
\end{eqnarray*}
\par 
The $\chi^2$--divergence between $f_Y^{(1)}$ and $f_Y^{(0)}$ 
is bounded as follows
\begin{multline*}
 \chi^2(f_Y^{(1)}, f_Y^{(0)}) = \int_0^\infty \frac{(f_Y^{(1)}(y)- f_Y^{(0)}(y))^2}{f_Y^{(0)}(y)} \rd y
 \\
 \;\;\;\leq c_6 \theta^2 h^2 \int_0^{\infty}  [1+2\ln^2(y/x_0)]\frac{1}{y} 
 \rho^2\big(\ln(y/x_0)\big)\rd y
 = c_6 \theta^2 h^2 \int_{-\infty}^\infty (1+2t^2)\rho^2 (t) \rd t.
\end{multline*}
By Parseval's identity, definition of $\psi$ and Assumption~[G1]
\begin{equation}\label{eq:chi-2}
 \int_{-\infty}^\infty \rho^2(t) \rd t = 
 \int_{-\infty}^\infty |\widetilde{g}(1+i\omega)|^2 |\widehat{\psi}(-\omega h)|^2 \rd \omega 
\leq   2 \int_{1/h}^{2/h} |\widetilde{g}(1+i\omega)|^2 \rd \omega  \leq c_7 h^{2\gamma-1}.
 \end{equation}
Moreover, using Assumptions~[G1] and~[G1$^\prime$]
\begin{multline*}
 \int_{-\infty}^\infty t^2 \rho^2(t)\rd t = \int_{-\infty}^\infty \Big|\frac{\rd}{\rd \omega} \widetilde{g}(1+i\omega) 
 \widehat{\psi}(-\omega h)\Big|^2\rd \omega
 \\
 \leq 2 \int_{-\infty}^\infty |\widetilde{g}^\prime (1+i\omega)|^2 |\widehat{\psi}(-\omega h)|^2 \rd \omega + 
 2 \int_{-\infty}^\infty |\widetilde{g} (1+i\omega)|^2 |\widehat{\psi}^\prime (-\omega h)|^2 h^2 \rd \omega
 \\
 \leq c_8 h^{2\gamma-1} + c_9 h^{2\gamma+1}.
\end{multline*}
Combining these bounds with (\ref{eq:chi-2}) for $h$ small enough we obtain 
\begin{eqnarray*}
 \chi^2(P^{(1)}, P^{(0)}) \leq c_9  \theta^2h^{2\gamma+1} = c_{10} 
 A^2 x_0^{2\beta+2} h^{2\beta+2\gamma+1}. 
\end{eqnarray*}
\par\medskip 
4$^0$. Now we complete the proof.
Let
\[
 h=h_*:= c_{11} x_0^{-\frac{2\beta+2}{2\beta+2\gamma+1}} (A^2n)^{-\frac{1}{2\beta+2\gamma+1}}.
\]
With this choice 
$\theta=c_2 A x_0^{\beta+1}h_*^\beta \leq \tfrac{1}{2}$ for $n$ large enough so 
$f_X^{(1)}\in \sH_{x_0,r}(A,\beta)$.  
We obtain $\chi^2(f_Y^{(1)}, f_Y^{(0)})\leq 1/n$ so that the hypotheses 
$f_X=f^{(0)}_X$ and $f_X=f^{(1)}_X$ 
are indistinguishable from the observations $Y_1,\ldots, Y_n$. Moreover,  
with this choice of the parameter $h$
\begin{multline*}
 \Big|f_X^{(1)}(x_0)- f_X^{(0)}(x_0)\Big| = 
 \theta |\psi_{h_*}(x_0)| = c_{12} Ax_0^{\beta+1}h_*^\beta x_0^{-1}|\psi(0)|
\\
 =
c_{13} 
A^{\frac{2\gamma+1}{2\beta+2\gamma+1}} 
x_0^{\frac{\beta(2\gamma-1)}{2\beta+2\gamma+1}} n^{-\frac{\beta}{2\beta+2\gamma+1}}~.
 \end{multline*}
This completes the proof of the theorem.
\epr
\subsection{Proof of Theorem~\ref{th:upper-bound-2}}
The bound on bias of $\hat{f}_{s,h}(x_0)$ given in Lemma~\ref{lem:bias} remains intact. 
We consider only the variance term. 
For 
\begin{eqnarray*}
  L_{s,h}(x,y) = 
  \frac{1}{2\pi x} \int_{-\infty}^{\infty} 
  \bigg(\frac{x}{y}\bigg)^{s+i\omega}\,  \frac{\widecheck{K}((s+i\omega)h)}{\widetilde{g}(1-s-i\omega)}
  \,\rd \omega
\end{eqnarray*}
we have 
\begin{multline*}
 \bE_{f_X} \big[L^2_{s,h}(x_0, Y_j)\big] 
\nonumber
 \\
 = 
 \frac{1}{4\pi^2 x_0^{2-2s}} \int_{-\infty}^\infty \int_{-\infty}^\infty x_0^{i(\omega-\mu)} 
 \widetilde{f}_Y(1-2s-i(\omega-\mu))\frac{\widecheck{K}((s+i\omega) h)}{\widetilde{g}(1-s-i\omega)}
 \cdot \overline{\frac{\widecheck{K}((s+i\mu) h)}{\widetilde{g}(1-s-i\mu)}}\rd \omega \rd \mu
 \nonumber
 \\*[2mm]
 \leq
 \frac{1}{4\pi^2x_0^{2-2s}} 
 \int_{-\infty}^\infty  |\widetilde{f}_Y(1-2s-i\mu)| \rd \mu 
 \int_{-\infty}^\infty
\bigg| \frac{
  \widecheck{K}((s+i\omega) h)}
 {\widetilde{g}(1-s-i\omega)}\bigg|^2\rd \omega.
 \end{multline*}
Since $f_X\in \sF_{\alpha, M}(A,\beta)$, 
$|\widetilde{f}_X(1-2s-i\mu)|\leq 1+M<\infty$ for
 all $\mu\in \bR$ and $-\alpha\leq s \leq 0$.
 For such $s$
 \begin{eqnarray*}
 \int_{-\infty}^\infty |\widetilde{f}_Y(1-2s -i\mu)|\rd \mu \leq (1+ M) \int_{-\infty}^\infty
 |\widetilde{g}(1-2s-i\mu)|\rd \mu \;\leq\; c_1 (1+M),
\end{eqnarray*}
 provided that $a<1-2s<b$.
 Setting $s=s_*=\max\{-\alpha, \tfrac{1}{2}(1-b)+\epsilon\}$ for any $\epsilon>0$ we obtain
 \begin{eqnarray}
 \bE_{f_X} \big[L^2_{s_*,h}(x_0, Y_j)\big] \leq 
 \frac{c_2(1+M)}{4\pi^2x_0^{2-2s_*}}  
 \int_{-\infty}^\infty
\bigg| \frac{
  \widecheck{K}((s_*+i\omega) h)}
 {\widetilde{g}(1-s_*-i\omega)}\bigg|^2\rd \omega.
\label{eq:sec-moment}
 \end{eqnarray}
\par
 Furthermore,  
\begin{eqnarray*}
 \widecheck{K}((s_*+i\omega)h)=
 \int_{-1}^1 K(x) e^{-s_*h x} e^{-i\omega h x}\rd x = \cF[v_{s_*,h}; \omega h]=
 \widehat{v}_{s_*,h}(\omega h),
\end{eqnarray*}
where $v_{s,h}(x):=K(x)e^{-shx}{\bf 1}_{[-1,1]}(x)$. 
Therefore 
\begin{multline}\label{eq:term-2}
 \int_{-\infty}^\infty \bigg| \frac{
  \widecheck{K}((s_*+i\omega) h)}
 {\widetilde{g}(1-s_*-i\omega)}\bigg|^2\rd \omega =
 \int_{-\infty}^\infty \bigg| \frac{
  \widehat{v}_{s_*,h}(-\omega h)}
 {\widetilde{g}(1-s_*-i\omega)}\bigg|^2\rd \omega
 \\
 \leq \frac{c_3}{h^{2\gamma+1}} \int_{-\infty}^\infty 
 |\widehat{v}_{s_*,h}(\omega)|^2 (1+ |\omega|^{2\gamma})\rd \omega,
\end{multline}
where $c_3$ may depend on $s_*$. In view of (\ref{eq:kernel-K-2}), $v_{s,h}$ is $q$ times 
continuously differentiable on its support,  and  
$v_{s,h}^{(q)}(x)= \sum_{j=0}^q  \tbinom{q}{j} K^{(j)}(x) (-sh)^{q-j} e^{-shx}$.
Therefore 
\begin{multline*}
 \big\|v_{s_*,h}^{(q)}\big\|_2 \leq  \sum_{j=0}^q \tbinom{q}{j} e^{2|s_*|h} 
 |s_*h|^{q-j} \bigg[\int_{-1}^1 |K^{(j)}(x)|^2\rd x\bigg]^{1/2} 
 \\
 \leq c_4 \max_{j=0,\ldots, q}
 \|K^{(j)}\|_2 \;\leq\; c_5C_K.
\end{multline*}
Taking into account that $q> \gamma+1$ and combining this 
inequality with (\ref{eq:term-2}) and (\ref{eq:sec-moment}) we obtain 
\begin{eqnarray*}
 {\rm var}_{f_X}\big\{\hat{f}_{s_*, h}(x_0)\} \leq c_6 (1+M) x_0^{-2+2s_*}h^{-2\gamma-1}n^{-1}.
\end{eqnarray*}
This bound together with the bound on the bias leads to the announced result.

\epr
\subsection{Proof of Theorem~\ref{th:upper-supersmooth}}
The proof goes along 
the same lines as the proof of Theorem~\ref{th:upper-bound-G1}.
In the proof below $c_1,c_2,\ldots$ stand for positive constants depending on $\beta$ and $\lambda$ only.
\par 
It is immediate to verify that 
\begin{equation}\label{eq:K-omega}
|\widehat{K}(\omega)|\leq c_1 \exp\{-\omega^{2\lambda}/2\lambda\},\;\;\;\forall \omega\in \bR. 
\end{equation}
This fact together with Assumption~[G2] guarantees that the estimator $\hat{f}_{h_*}(x_0)$
is well--defined.
In addition, by \cite[Chapter~IV, \S~7]{fedoryuk1987asymptotics} as $|t|\to\infty$
\begin{multline*}
 w(t)=2\sqrt{\tfrac{1}{2\lambda-1}} |t|^{-(\lambda-1)/(2\lambda-1)} 
 \exp\Big\{-\Big(\tfrac{2\lambda-1}{2\lambda}\Big) 
 \sin \Big(\tfrac{\pi}{2(2\lambda-1)} \Big)|t|^{2\lambda/(2\lambda-1)}\Big\} 
\\*[2mm]
 \times \Big[\cos \bigg(\tfrac{2\lambda-1}{2\lambda} |t|^{2\lambda/(2\lambda-1)}
 \cos\Big(\frac{\pi}{2(2\lambda-1)}\Big)\bigg) + O
 \Big(|t|^{-2\lambda/(2\lambda-1)}\Big)\Big].
\end{multline*}
Therefore, it follows from (\ref{eq:K-supersmooth}) that for large $|t|$ one has
\begin{eqnarray}\label{eq:K-bound-supsmooth}
 |K(t)|
 \leq  c_2 |t|^{-(\lambda-1)/(2\lambda-1)} 
 \;\exp\Big\{-c_3 |t|^{2\lambda/(2\lambda-1)}\Big\}.
 \end{eqnarray}
\par 
First we bound the bias of the  estimator $\hat{f}_{h_*}(x_0)$. To that end we note that the proof of 
Lemma~\ref{lem:bias} applies verbatim; the only difference is that now 
the integration in   (\ref{eq:K-w}) is over the whole
real line because $K$ is not compactly supported.  
 However, since $K$  is a bounded function and in view of (\ref{eq:K-bound-supsmooth})
 we have 
\begin{eqnarray*}
 \int_{-\infty}^\infty |t|^{\ell+1} e^{\ell |th|} |K(t)|\rd t \leq 
 \int_{-\infty}^\infty |t|^{\ell+\frac{\lambda}{2\lambda-1}} \exp\Big\{ \ell |t| - c_3 |t|^{-\frac{2\lambda}{2\lambda-1}}
 \Big\}
 \rd t
 \leq 
 c_4.
\end{eqnarray*}
This inequality and reasoning of the proof of Lemma~\ref{lem:bias} 
yield
\[
\sup_{f\in \sH_{x,r}(A,\beta)} \Big|\int_{-\infty}^\infty
K_{h}(x,y)f(y)\rd y - f(x)\Big| \leq c_5 
A \Big[h^\beta |x|^\beta + h^{\ell+1}\sum_{k=0}^\ell |x|^k\Big]. 
\]
\par 
To bound the variance we follow the lines of the proof of Theorem~\ref{th:upper-bound-G1}. In particular, 
in view of (\ref{eq:K-omega}) and Assumption~[G2] by straightforward algebra we have for small enough $h$ 
\begin{multline*}
 \bE_{f_X}\big[L_h^2(x_0, Y_j)\big] \leq \frac{c_6}{x_0^2} \int_{-\infty}^\infty 
 \frac{|\widehat{K}(\omega h)|^2}
 {|\widetilde{g}(1-i\omega)|^2}\rd \omega 
 \\
 \leq \frac{c_7}{x_0^{2}} \bigg\{1+ \int_{\omega_0\leq |\omega|\leq \omega_1} + \int_{|\omega|>\omega_1}
 |\omega|^{-2\nu}\exp\{|\omega|\gamma -|\omega h|^{2\lambda} /\lambda\} \rd \omega \bigg\}
 \\
\leq \frac{c_8}{x_0^2}
 \exp\Big\{c_9\big(\gamma h^{-1}\big)^{2\lambda/(2\lambda-1)}\Big\},
\end{multline*}
where we set $\omega_1:=(\lambda\gamma)^{1/(2\lambda-1)} h^{-2\lambda/(2\lambda-1)}$. Then 
the result of the theorem follows from balancing the bounds in the two previous display formulas.
\epr 

\subsection{Proof of Theorem~\ref{thm:upper_zero}}
In the proof below $c_1, c_2, \ldots$ stand for positive constants; they can depend on parameters appearing in
assumptions [G3] and [G4] and on  parameter $\beta$ only.
The proof proceeds in steps.
\par\medskip
1$^0$. First we show that 
under the premise of the theorem the estimator $\hat{f}_{s_*,h}(0)$ is well defined. 
It follows from the definition of function $\psi_s(x)$ that
\begin{multline*}
 \widetilde{\psi}_s(s+i\omega)= \frac{1}{\sqrt{2\pi}}e^{-\frac{1}{2}(1-s)^2}
 \int_0^\infty x^{s+i\omega-1} x^{-s}\exp\{-\tfrac{1}{2}[\ln x]^2\}\rd x
 \\
 = \frac{1}{\sqrt{2\pi}}e^{-\frac{1}{2}(1-s)^2}
 e^{-\frac{1}{2}\omega^2};
\end{multline*}
therefore 
\begin{eqnarray*}
 \int_0^\infty \psi_s\Big(\frac{x}{j}\Big) x^{s+i\omega-1}\rd x= j^{s+i\omega} 
 \widetilde{\psi}_s(s+i\omega)= \frac{j^{s+i\omega}}{\sqrt{2\pi}}e^{-\frac{1}{2}(1-s)^2}
 e^{-\frac{1}{2}\omega^2},
\end{eqnarray*}
and 
\begin{eqnarray*}
 \widetilde{K}_s(s+i\omega) = \frac{1}{\sqrt{2\pi}}e^{-\frac{1}{2}(1-s)^2}
 e^{-\omega^2/2}
 \sum_{j=1}^{m+1} \tbinom{m+1}{j} (-1)^{j+1} j^{s-1+i\omega}. 
\end{eqnarray*}
The last expression implies that 
\begin{equation}\label{eq:tilde-K}
 |\widetilde{K}_s(s+i\omega)| \leq c_1 m^s e^{-\frac{1}{2}(1-s)^2}
 e^{-\omega^2/2},
\end{equation}
where $c_1$ depends on $m$ only.
Next we observe that Assumption~[G3] implies $1-s_*=1-\frac{1}{2}(1-p)\in \Omega_g$, so that 
$\widetilde{g}(1-s_*+i\omega)$ is well defined. Then in view of (\ref{eq:tilde-K}) and 
condition (\ref{eq:G4-1}) of
Assumption~[G4], $\widetilde{K}_{s_*}(s_*+i\cdot)/ \widetilde{g}(1-s_*-i\cdot)\in \bL_1(\bR)\cap \bL_2(\bR)$
so that $\hat{f}_{s_*,h}(0)$ is well defined.
\par\medskip 
2$^0$. Our next step is to prove the following statement about local behavior of the density $f_Y$ near the 
origin. This result is instrumental in establishing an upper bound on the variance term. 
\begin{lemma}\label{lem:f-y-nearzero}
 Let Assumption~[G3] hold,  and assume that $f_X(t)\leq M$, $\forall t$.
 \begin{itemize}
  \item[{\rm (i)}] If $p=0$ then for all $y\leq \delta$ 
  \[
   f_Y(y)\leq  C_1 (1+M)|\ln y|^{q+1} + M\delta^{-1},
  \]
  where $C_1$ depends on $q$ only.
\item[{\rm (ii)}] If $p\in (0,1)$ then for all $y\leq \delta$
\[
 f_Y(y) \leq C_2 \big(1+Mp^{-1}\big) y^{-p} |\ln y|^q +M\delta^{-1},
\]
where $C_2$ depends on $q$ only.
\end{itemize}
\end{lemma} 
The proof of the lemma is given in Section~\ref{sec:auxiliary}.
%
%
\par\medskip 
3$^0$. Now we are ready to establish an upper bound on the variance term. 
Define 
\[
 \rho_s(x):=\frac{1}{2\pi}\int_{-\infty}^\infty e^{-i\omega x}
 \frac{\widetilde{K}_s(s+i\omega)}{\widetilde{g}(1-s-i\omega)} \rd \omega.
\]
With this notation $L_{s,h}(y)=h^{s-1} y^{-s} \rho_s(\ln (y/h))$
[cf. (\ref{eq:Lsh})],
and therefore 
\[
 \bE_{f_X} \big[L^2_{s,h}(Y)\big]= \frac{1}{h^{2-2s}}\int_0^\infty \frac{\rho_s^2(\ln (y/h))}{y^{2s}} f_Y(y) \rd y.
\]
Now we  bound the last integral which can be  written as a sum $J_1+J_2$, where 
 \[
 J_1:=\frac{1}{h^{2-2s}}\int_0^{\delta} \frac{\rho_s^2(\ln (y/h))}{y^{2s}} f_Y(y) \rd y,\;
 J_2:=\frac{1}{h^{2-2s}}\int_{\delta}^\infty \frac{\rho_s^2(\ln (y/h))}{y^{2s}} f_Y(y) \rd y.
 \]
 Using  Lemma~\ref{lem:f-y-nearzero} for $p=0$ and $s=s_*=\frac{1}{2}$ by straightforward algebra we obtain
\begin{eqnarray*}
 J_1 &\leq& \frac{c_1M}{h} \int_0^\delta y^{-1} \rho_{s_*}(\ln(y/h))\big[|\ln y|^{q+1}+ \delta^{-1}\big]\rd y
 \\
 &\leq & \frac{c_2M}{h} \bigg( |\ln h|^{q+1} \int_{-\infty}^\infty \rho_{s_*}^2(t) 
 \rd t + 
 \int_{-\infty}^\infty \rho_{s_*}^2 (t) |t|^{q+1}\rd t\bigg)
\leq  c_3 M h^{-1} |\ln h|^{q+1}, 
 \end{eqnarray*}
where the last inequality follows from condition~[G4] [cf. (\ref{eq:G4-1}) and (\ref{eq:G-4-2})].
If $p\in (0,1)$ then using Lemma~\ref{lem:f-y-nearzero}
for $s=s_*=\frac{1}{2}(1-p)$ we have similarly
\begin{eqnarray*}
 J_1 &\leq& 
 \frac{c_4M}{h^{2-2s_*}}\int_0^{\delta_0} \frac{1}{y} \rho_{s_*}^{2}(\ln (y/h)) 
 \frac{y^{-p}|\ln y|^{q+\kappa}}{y^{2s_*-1}}\rd y 
 \\
 &\leq& \frac{c_5M}{h^{1+p}} \bigg( |\ln h|^q
 \int_{-\infty}^\infty \rho_{s_*}^2(t) \rd t + \int_{-\infty}^\infty \rho_{s_*}^2(t) |t|^{q}\rd t
 \bigg) \leq  c_6M h^{-1-p} [\ln (1/h)]^{q}.
\end{eqnarray*}
Combining the last two upper bounds on $J_1$ in cases $p=0$ and $p\in (0,1)$ we can write 
\[
 J_1 \leq  c_7M h^{-1-p} [\ln (1/h)]^{q+\kappa},
\]
where $\kappa$ is defined in (\ref{eq:kappa}).
\par 
In order to bound $J_2$ we note that (\ref{eq:G4-1}) implies 
$|\rho_{s_*}(x)|\leq c_8<\infty$, $\forall x$; therefore 
\[
 J_2 \leq c_8^2h^{-1-p} \int_{\delta_0}^\infty y^{p-1} f_Y(y) \rd y \leq c_9 h^{-1-p}.
\]
Combining the bounds on $J_1$ and $J_2$  we obtain
\[
 \bE_{f_X} \big[L^2_{{s_*},h}(Y)\big] \leq c_{10}  M h^{-1-p} [\ln (1/h)]^{q+\kappa}.
\]
\par\medskip 
4$^0$. We proceed with bounding the bias of 
 $\hat{f}_{s_*,h}(0)$. By construction of $K_{s_*,h}(x)$ we have 
\begin{eqnarray*}
 \int_0^\infty (1/h) K_{s_*}(x/h)[f_X(x)-f_X(0)]\rd x =\int_0^\infty K_{s_*}(u)[f_X(uh)-f_X(0)]\rd x
 \\
 = \int_0^{r/h} K_{s_*}(u) \Big[\sum_{j=1}^{\ell -1} \frac{1}{j!} f_X^{(j)}(0)(uh)^j + \frac{1}{\ell!}
 f_X^{(\ell)} (\xi uh) (uh)^{\ell}\Big]\rd u 
 \\
 + \int_{r/h}^\infty K_{s_*}(u)[f_X(uh)-f_X(0)]\rd u.
\end{eqnarray*}
Since $f_X(x)\leq M$, $\forall x$, by (\ref{eq:psi-sigma}) and (\ref{eq:K-sigma})
\begin{multline*}
 \bigg|\int_{r/h}^\infty K_{s_*}[f_X(uh)-f_X(0)]\rd u\bigg|\leq 2M \sum_{j=1}^{m+1}
 \tbinom{m+1}{j} \int_{r/h}^\infty \psi_{s_*}(x)\rd x 
 \\
 \leq c_1M\exp\{-c_2 [\ln(r/h)]^2\}. 
\end{multline*}
Furthermore, it is readily verified that for small enough $h$
\[
 \bigg|\int_0^{r/h} K_{s_*}(u) u^j\rd u\bigg|=\bigg|\int_{r/h}^\infty K_{s_*}(u)u^j\rd u\bigg|\leq 
 c_{3} \exp\big\{-c_{4} [\ln\big(\tfrac{r}{mh}\big)]^2\big\}.
\]
Using these facts we finally obtain that 
\begin{multline*}
 \bigg|\int_0^\infty (1/h) K_{s_*}(x/h)[f_X(x)-f_X(0)]\rd x\bigg|
 \\
 \leq c_{5}Ah^\beta+ c_{6}(1+M)
 \exp\big\{-c_{7} \big[\ln \big(\tfrac{r}{mh}\big)\big]^2\big\}.
\end{multline*}
We complete the proof by noting that 
the choice $h=h_*$ indicated in the statement of the theorem provides 
a balance for the bounds on the bias and 
on the variance.
\epr

\subsection{Proof of Theorem~\ref{th:lower-bound-zero}}
The proof is based on the standard technique for proving lower bounds (see 
\cite[Chapter~2]{tsybakov2009introduction}). Throughout the proof constants $c_0, c_1,\ldots$ 
may depend only on $\beta$ and parameters appearing in Assumptions~[G3] and [G5].
\par\medskip 
1$^0$. Let $M_0=\pi M/4$, and without loss of generality assume that $M_0\geq 1$. Let  
\[
f_{X}^{(0)}(x):=\frac{2M_0}{\pi(1+M_0x)(1+\ln^{2}(1+M_0x))},\quad x\geq 0.
\]
It is evident that $f_X^{(0)} (x) \leq M/2$, $\forall x$ and  $f_X^{(0)}\in \bar{\sH}_r(A, \beta, M)$ provided 
that $A$ is large enough.
\par 
For  $h>0$ define
\[
f_{X}^{(1)}(x)=f_{X}^{(0)}(x)+c_0 A h^{\beta}\varphi\left(x/h\right),\;\;\;
\varphi(x):=(1-x)e^{-x},\;\;\;x\geq 0.
\]
In what follows parameter $h$ will be chosen going to zero as $n\to\infty$; in the subsequent proof 
we use this fact.
It is evident that  function $f_X^{(1)}$ is a probability density, and 
under appropriate choice of constant $c_0$ and for $h$ small enough
it belongs 
to $\bar{\sH}_{r}(A,\beta,M)$. We note also that 
\begin{equation}\label{eq:varphi-tilde}
 \widetilde{\varphi}(z)=\int_0^\infty x^{z-1} \varphi(x)\rd x= \Gamma(z)-\Gamma(z+1),\;\; z\in \Omega_\varphi=
 \{z: {\rm Re}(z)>0\}.
\end{equation}
\par 
Our current goal is to bound the $\chi^2$--divergence between the corresponding densities 
of observations $f_Y^{(0)}$ and $f_Y^{(1)}$.
For any $s$ such that $\{z: {\rm Re}(z)=s\} \subseteq  \Omega_g\cap \Omega_\varphi$
we have 
\begin{align*}
 f_Y^{(1)}(y)-&f_Y^{(0)}(y) = 
 c_0A h^\beta \int_0^\infty \frac{1}{x}\varphi\Big(\frac{y}{hx}\Big)g(x)\rd x
 \\
&= \frac{c_0 A h^\beta}{2\pi i}\int_{s-i\infty}^{s+i\infty} \Big(\frac{h}{y}\Big)^z 
\widetilde{\varphi}(z)\widetilde{g}(z)\rd z
\\
&=
\frac{c_0A h^{\beta+s}}{2\pi y^s} \int_{-\infty}^\infty 
\Big(\frac{h}{y}\Big)^{i\omega} \widetilde{\varphi}(s+i\omega) \widetilde{g}(s+i\omega)\rd \omega =
c_0A h^{\beta+s} y^{-s} \rho_s(\ln(y/h)),
 \end{align*}
where we have used the Mellin transform inversion formula, and we have denoted 
\begin{equation}\label{eq:rho-s}
 \rho_s(t):=\frac{1}{2\pi}\int_{-\infty}^\infty 
 e^{-i\omega t}\widetilde{\varphi}(s+i\omega)\widetilde{g}(s+i\omega)\rd \omega = 
 \frac{e^{st}}{2\pi}\int_{s-i\infty}^{s+i\infty} e^{-zt} \widetilde{\varphi}(z)\widetilde{g}(z)
 \rd z.
\end{equation}
Thus 
\begin{equation}\label{eq:chi-22}
 \chi^2(f_Y^{(0)}, f_Y^{(1)}) = c_0^2A^2 h^{2(\beta+s)}\int_0^\infty \frac{y^{-1}\rho_s^2(\ln (y/h))}
 {y^{2s-1} f_Y^{(0)}(y)} \rd y,
\end{equation}
and 
now we will bound the integral on the right hand side under a particular choice of parameter 
$s$.
\par\medskip 
2$^0$.
Let $s=s_*:=\frac{1}{2}(p+1)$. 
Note that
by  the upper bound in 
(\ref{eq:g-near-zero})
and by definition of $s_*$ 
\[
 |\widetilde{g}(s_*+i\omega)| \leq  \int_0^\infty x^{s_*-1} g(x)\rd x =
 \int_0^\infty x^{(p-1)/2} g(x)\rd x \leq c_1;
\]
thus $\{z: {\rm Re}(z)=s_*\}\subseteq \Omega_{g}$.
Let $\nu:= \frac{1}{2}(1-p)+\epsilon$, where $\epsilon$
is given  in Assumption~[G5].
Then
$s_*+\nu =1+\epsilon$, and according to 
Assumptions~[G3] and [G5],  function 
$e^{-zt}\widetilde{g}(z) \widetilde{\varphi}(z)$ is analytic in 
$\{z: s_*\leq {\rm Re}(z)\leq s_*+\nu\}$.
Therefore 
the line of integration in the last integral on the right hand side of 
(\ref{eq:rho-s}) 
can be replaced by  
$\{z: {\rm Re}(z)=s_*+ \nu\}$. This   yields 
\begin{eqnarray*}
 \rho_{s_*}(t)&=&\frac{e^{s_*t}}{2\pi} \int_{s_*+\nu-i\infty}^{s_*+\nu +i\infty} e^{-zt} 
 \widetilde{g}(z) \widetilde{\varphi}(z)
 \rd z 
 \\
 &=& 
 \frac{e^{-\nu t}}{2\pi} \int_{-\infty}^\infty e^{-i\omega t} \widetilde{g}(1+\epsilon +i\omega) 
 \widetilde{\varphi}(1+\epsilon +i\omega)
 \rd \omega.
\end{eqnarray*}
 Then it follows from Assumption~[G5] that   
\begin{equation}\label{eq:rho-bound}
 |\rho_{s_*}(t)| \leq c_2 e^{-\nu t} \int_{-\infty}^\infty |
 \widetilde{\varphi}(1+\epsilon +i\omega)| \rd \omega  \leq c_3 e^{-\nu t},
\end{equation}
where the last inequality follows from 
(\ref{eq:varphi-tilde}) and   bounds on the Gamma function  as presented in (\ref{gamma_asymp}) in Example~4.
\par\medskip 
3$^0$. 
Now we derive lower bounds on $f_Y^{(0)}(y)$.
Note that  $f_X^{(0)}(x)= M_0 \bar{f}_X^{(0)}(M_0x)$ where 
\[
 \bar{f}_X^{(0)}(x) :=\frac{2}{\pi(1+x)(1+\ln^{2}(1+x))},\quad x\geq 0.
\]
Therefore $f_Y^{(0)}(y) = M_0 \bar{f}_Y^{(0)}(M_0y)$,  $\bar{f}_Y^{(0)}(y):= [\bar{f}_X^{(0)} \star g](y)$ and  
the lower bounds on $f_Y^{(0)}(y)$ can be obtained in an evident way from the corresponding bounds on
$\bar{f}_Y^{(0)}(y)$.
\par 
First we note that the lower bound in (\ref{eq:g-near-zero}) and  
the arguments  as in the proof of (\ref{eq:ypp}) in Lemma~\ref{lem:f-y-nearzero}, 
yield for all $y < \delta/2$
\begin{equation}\label{eq:y-delta}
 \int_{y}^\delta [g(t)/t]\rd t \geq c_4 y^{-p}|\ln y|^{q+\kappa},
\end{equation}
where $\kappa$ is defined in (\ref{eq:kappa}).
In view of (\ref{eq:y-delta}) for $y< \delta/2$ 
\begin{eqnarray}
\bar{f}_{Y}^{(0)}(y)  &\geq& \int_{y}^{\delta}\frac{2g(x)}{\pi x(1+y/x)(1+\ln^{2}(1+y/x))}\,\rd x
\nonumber
\\
&
\geq& \frac{1}{\pi (1+\ln^2(2))}
\int_y^\delta \frac{g(x)}{x}\,\rd x \geq c_5y^{-p}|\ln y|^{q+\kappa}.
\nonumber
\end{eqnarray}
Thus, 
\begin{equation}
\label{eq:f-y-0}
 f_Y^{(0)}(y)\geq c_5 M_0^{1-p} y^{-p} |\ln (M_0y)|^{q+\kappa}, \;\;\;\forall y<\delta/(2M_0).
\end{equation}
On the other hand,
for any $y$ we have 
\begin{eqnarray}
\bar{f}_{Y}^{(0)}(y)  
&=& \int_{0}^{\infty}\frac{2g(x)}{\pi x(1+y/x)(1+\ln^{2}(1+y/x))}\,\rd x
\nonumber
\\
&\geq& \int_{0}^{1}\frac{2g(x)}{\pi (x+y)(1+2\ln^{2}(x+y)+2\ln^{2}(x))}\,\rd x 
\nonumber
\\
 &\geq& \frac{2}{\pi (1+y)(1+2\ln^{2}(1+y))}\int_{0}^{1}\frac{g(x)}{1+2\ln^{2}(x)}\,\rd x
\nonumber
 \\
 &\geq& \frac{c_{6}}{(1+y)(1+2\ln^{2}(1+y))},
\nonumber
 \end{eqnarray}
 so that 
 \begin{equation}\label{eq:f_y>>}
 f_Y^{(0)}(y) \geq \frac{c_{6}M_0}{(1+M_0y)(1+2\ln^{2}(1+M_0y))},\;\;\;\forall y.
 \end{equation}

\par\medskip 
4$^0$. Now we bound from above the integral on the right hand side of (\ref{eq:chi-22}).
\par 
Let $\xi\in (h, \delta/(2M_0))$ be a parameter that will be specified later; then 
we  can write the integral on the right hand side of (\ref{eq:chi-22}) in the following form
\begin{align}\label{eq:I1+I2}
& \int_{0}^{\infty}\frac{y^{-1}\rho_s^{2}\left(\ln(y/h)\right)}
{y^{2s-1}f_{Y}^{(0)}(y)}\,\rd y 
 \nonumber
 \\
&\;\; = \int_{0}^{\xi}
\frac{y^{-1}\rho_s^{2}\left(\ln(y/h)\right)}{y^{2s-1}f_{Y}^{(0)}(y)}\,\rd y+
\int_{\xi}^\infty\frac{y^{-1}\rho_s^{2}\left(\ln(y/h)\right)}{y^{2s-1}f_{Y}^{(0)}(y)}\,\rd y
 =: I_{1}+I_{2}.
\end{align}
Our current goal is to bound $I_1$ and $I_2$ when $s=s_*=\frac{1}{2}(p+1)$. 
\par 
Using  (\ref{eq:f-y-0}) we obtain
\begin{align}
 I_1 \leq \frac{c_6}{M_0^{1-p}} \int_0^{\xi}\frac{y^{-1}\rho_{s_*}^2(\ln (y/h))}{|\ln(M_0y)|^{q+\kappa}}\rd y 
 \leq  \frac{c_6|\ln (M_0\xi)|^{-q-\kappa}}{M_0^{1-p}} \int_{-\infty}^{\ln (\xi/h)}
 \rho_{s_*}^2(t)
 \rd t 
\label{eq:I-11}
\end{align}
It follows from (\ref{eq:rho-s}) that 
\begin{eqnarray*}
\int_{-\infty}^\infty \rho^2_{s_*}(t)\rd t &=& \frac{1}{2\pi}
 \int_{-\infty}^\infty |\widetilde{\varphi}(s_*+i\omega)|^2 |\widetilde{g}(s_*+i\omega)|^2\rd \omega
 \nonumber
 \\
 &\leq& c_7 \int_{-\infty}^\infty |\widetilde{\varphi}(s_*+i\omega)|^2\rd \omega 
 = c_7\int_0^\infty x^{2s_*-1} \varphi^2(x)\rd x \leq c_8,
 \end{eqnarray*}
where the equality in the last line follows from the Parseval identity \eqref{eq:parceval-1}, and the 
last inequality is by definition of $\varphi$.
This inequality together with (\ref{eq:I-11}) leads to 
\begin{equation}\label{eq:I-111}
 I_1 \leq c_9 M_0^{-1+p} |\ln(M_0\xi)|^{-q-\kappa}. 
\end{equation}
\par 
Now consider the integral $I_2$ on the right hand side of (\ref{eq:I1+I2}).
Using (\ref{eq:f_y>>}) we write (remind that $2s_*-1=p$)
\begin{align}
 I_2 &\leq  \frac{c_{10}}{M_0}\int_{\xi}^\infty y^{-p-1} \rho_{s_*}^2(\ln (y/h)) (1+M_0y)[1+\ln^2(1+M_0y)]\rd y
\nonumber
 \\
&= \frac{c_{10}}{M_0} \bigg\{ 
\int_{\xi}^\infty y^{-p-1}[1+\ln^2(1+M_0y)] \rho_{s_*}^2(\ln (y/h)) \rd y 
\nonumber
\\
 &+\; M_0
\int_{\xi}^\infty y^{-p}[1+\ln^2(1+M_0y)] \rho_{s_*}^2(\ln (y/h)) \rd y
\bigg\} =: \frac{c_{10}}{M_0} \{ I_2^{(1)} + I_2^{(2)}\}.
 \label{eq:I-22}
 \end{align}
Applying (\ref{eq:rho-bound}), using a simple inequality 
\(\ln(1+x)\leq\ln 2+ \left|\ln x\right|,\)  \(x\geq 0,\) and assuming that $h$ is small so that 
$M_0h\leq 1$ we derive 
\begin{eqnarray}
 I_2^{(1)} &=& h^{-p} \int_{\ln (\xi/h)}^\infty e^{-pt} \rho_{s_*}^2(t) \big[1+\ln^2(1+M_0he^t)\big]\rd t
\nonumber
 \\
&\leq&  c_{11} h^{-p}  \int_{\ln (\xi/h)}^\infty e^{-(p+\nu)t} e^{-\nu t}  (1+t^2)\rd t
\nonumber
\\
&\leq& c_{12}  h^{\nu} \xi^{-p-\nu}  
\int_0^\infty e^{-\frac{1}{2}(1-p)t} (1+t^2) \rd t 
\leq c_{13} h^{\nu} \xi^{-p-\nu},
 \label{eq:I-2-1}
 \end{eqnarray}
 and similarly
 \begin{align}
  I_2^{(2)} &\leq c_{14} M_0 h^{-p+1} \int_{\ln (\xi/h)}^\infty e^{t(1-p)} e^{-2\nu t}\big[1+ t^2\big]\rd t
  \leq 
 c_{15}M_0 h^{-p+1},  
\label{eq:I-2-2}
  \end{align}
where  we have used that $\nu=\frac{1}{2}(1-p)+\epsilon$. 
  Combining inequalities (\ref{eq:I-2-2}), (\ref{eq:I-2-1}), (\ref{eq:I-22}) and (\ref{eq:I-111})
we conclude that for small enough $h$ and  for $\xi\in (h,\delta/(2M_0))$ 
one has
\[
 \int_{0}^{\infty}\frac{y^{-1}\rho_{s_*}^{2}\left(\ln(y/h)\right)}
{y^{2s-1}f_{Y}^{(0)}(y)}\,\rd y  \leq \frac{c_{16}}{M_0} 
\Big\{ M_0^p[\ln (M_0/\xi)]^{-q-\kappa} + h^\nu \xi^{-p-\nu} +  M_0h^{-p+1}\Big\}.
 \]
\par 
 Let $\nu_0\in (0,\nu)$; then we set $\xi=h^{(\nu-\nu_0)/(p+\nu)}$. 
First, we note that with this choice $\xi\geq h$ as required. 
Second, it is immediately verified that  
the second term in the figure brackets on the right hand side of the previous display formula is bounded above by 
$h^{\nu_0}$, 
and the first term is dominant 
as $h\to 0$. 
Combining this result with (\ref{eq:chi-22}) we conclude that for $h$ small enough
 \begin{equation*}
 \chi^2(f_Y^{(0)}, f_Y^{(1)}) = c_{16}A^2 M^{-1+p} h^{2(\beta+s_*)}
 [\ln (1/h)]^{-q-\kappa},
\end{equation*}
where we took into account that $M_0=\pi M/4$.
\par\medskip 
5$^0$. Now we complete the proof of the theorem. Let 
\[
 h=h_*=\big[ c_{17}A^{-2} M^{1-p} (\ln n)^{q+\kappa} n^{-1} \big]^{1/(2\beta+1+p)}.
\]
With this choice and appropriately small constant $c_{16}$ the 
$\chi^2$--divergence
$\chi^2(f_Y^{(0)}, f_Y^{(1)})$ is less than $1/n$, and the hypotheses $f_X=f_X^{(0)}$ and $f_X=f_X^{(1)}$ cannot be 
distinguished from the observations. Under these circumstances 
\[
|f_X^{(0)}(0)-f_X^{(1)}(0)|= c_1 Ah_*^\beta = c_{18}A^{\frac{p+1}{2\beta+1+p}}
\big[ M^{1-p} (\ln n)^{q+\kappa} n^{-1} \big]^{\frac{\beta}{2\beta+1+p}}.
\]
This completes the proof.
\epr

\section{Proofs of auxiliary results}\label{sec:auxiliary}

\subsection{Proof of Lemma~\ref{lem:identifiability}}
Considering the integral (\ref{eq:f_Y}) for $y\geq 0$ and $y<0$ and using notation (\ref{eq:u+u-}) we obtain 
\begin{eqnarray}
f^+_Y(y)
&=&
\int_0^\infty \frac{1}{x} f^+_X(y/x) g^+(x)\rd x \;-\; \int_0^\infty \frac{1}{x}
f^-_X(y/x) g^-(x)\rd x 
\label{eq:fY+}\\*[2mm]
f_Y^-(y) &=&  - \int_0^\infty \frac{1}{x} f_X^+(y/x)g^-(x)\rd x + \int_0^\infty 
\frac{1}{x}f_X^-(y/x) g^+(x)\rd x.
\label{eq:fY-}
\end{eqnarray}
Applying the Mellin transform to the both sides of
(\ref{eq:fY+})--(\ref{eq:fY-}), we have
\begin{equation}\label{eq:system}
\begin{array}{l}
 \widetilde{f}_Y^+(z) = \widetilde{f}^+_X(z) \widetilde{g}^+(z) - \widetilde{f}_X^-(z) \widetilde{g}^-(z),\;\;\;
 \\*[2mm]
 \widetilde{f}_Y^-(z) = -\widetilde{f}^+_X(z) \widetilde{g}^-(z) + \widetilde{f}_X^-(z) \widetilde{g}^+(z).
\end{array}
\end{equation}
Note that the line $\{z: {\rm Re}(z)=1\}$ is in the strip of analyticity of $\widetilde{f}^\pm_X$
and $\widetilde{g}^\pm$ because $f_X$ and $g$ are probability densities. 
Thus the Mellin transforms 
in (\ref{eq:system}) are well--defined
in an infinite strip containing the line $\{z:{\rm Re}(z)=1\}$. 
\par 
The system of equations (\ref{eq:system}) has a
unique solution $(\widetilde{f}_X^+(z), \widetilde{f}_X^-(z))$
if and only~if 
\[
 \bigg|{\rm det}\left[ \begin{array}{cc}
        \widetilde{g}^+(z) & - \widetilde{g}^-(z)\\
        -\widetilde{g}^-(z) & \widetilde{g}^+(z)
       \end{array}\right]
       \bigg| = \big|[\widetilde{g}^+(z)]^2 - [\widetilde{g}^-(z)]^2\big|\ne 0.
\]
Under this condition, with $\widetilde{f}^+_X(z)$ and $\widetilde{f}_X^-(z)$ satisfying
(\ref{eq:system}) in the common region of analyticity containing the line $\{z: {\rm Re}(z)=1\}$, 
functions $f_X^+$ and $f_X^-$ are uniquely determined by the inversion
formula
%
\[
 f_X^\pm(x)=\frac{1}{2\pi}\int_{-\infty}^{\infty} x^{-(1+iv)} \widetilde{f}_X^\pm (1+iv) \rd v.
\]
Therefore the necessary and sufficient 
conditions for identifiability are 
\begin{equation}\label{eq:g+neg-}
 \widetilde{g}^+(z)-\widetilde{g}^-(z)={\int_0^\infty x^{z-1} [g(x)-g(-x)] \rd x}\ne 0 ,\;\;\;\widetilde{g}^+(z)+\widetilde{g}^-(z)\ne 0
\end{equation}
for almost all $z$ in the common strip of analyticity of $\widetilde{g}^+$ and $\widetilde{g}^-$.
Note that $\widetilde{g}^+(z)+\widetilde{g}^-(z)$ is  an analytic function; 
therefore the second condition in (\ref{eq:g+neg-})
holds for any density $g$. Then the statement of the lemma follows from 
the uniqueness property of the Mellin transform.
\epr

\subsection{Proof of Lemma~\ref{lem:L}}
By (\ref{eq:f_Y}) we have 
\begin{eqnarray*}
\int_{-\infty}^\infty L_{s,h}(x,y) f_Y(y) \rd y
=\int_{-\infty}^\infty \bigg[\int_{-\infty}^\infty L_{s,h}(x, ty)g(t)\rd t\bigg] f_X(y)\rd y;
\end{eqnarray*}
therefore, in order to prove 
(\ref{eq:K-L})
it suffices to show that $L_{s,h}(\cdot, \cdot)$ solves the equation 
\begin{equation}\label{eq:equation}
 \int_{-\infty}^\infty L_{s,h}(x,ty)g(t)\rd t = K_h(x,y).
\end{equation}
To this end,  we will show that for any fixed $x$ the one--sided Mellin transforms 
of expressions on the both sides of (\ref{eq:equation})
coincide in a common  vertical strip of the complex plane.
This will imply the lemma statement.
\par
It follows from (\ref{eq:K-kernel}) that for $x>0$
\begin{eqnarray} \label{eq:K-Mellin-1}
 \int_0^\infty y^{z-1} K_h(x,y)\rd y =  x^{z-1}\int_{-\infty}^\infty K(t) e^{thz}\rd t= x^{z-1}\widecheck{K}(zh),
 \end{eqnarray}
 and for $x<0$
 \begin{eqnarray}
 \int_{-\infty}^0 (-y)^{z-1}  K_h(x,y)\rd y =  (-x)^{z-1}\int_{-\infty}^\infty K(t) e^{thz}\rd t
 = 
 (-x)^{z-1} \widecheck{K}(zh).
 \label{eq:K-Mellin-2}
 \end{eqnarray}
Let
\[
L_{s,h}^+(\cdot,y):=\left\{\begin{array}{ll} 
                        L_{s,h}(\cdot,y), & y>0\\
                        0, & y<0,
                       \end{array}
                       \right. 
\;\;L_{s,h}^-(\cdot,y):=\left\{\begin{array}{ll} 
                            L_{s,h}(\cdot, -y), & y>0\\
                            0, & y<0.
                           \end{array}
 \right.
\]
Remind that with this notation,
$L_{s,h}(\cdot, y) = L_{s,h}^+(\cdot, y)$ for $y\geq 0$ and  
$L_{s,h}(\cdot, y)=L_{s,h}^-(\cdot,-y)$ for $y<0$.
Integrating the left hand side of (\ref{eq:equation}) we obtain 
\begin{align}
 &\int_0^\infty y^{z-1} \int_{-\infty}^\infty L_{s,h}(x,ty) g(t) \rd t \rd y 
 \nonumber
 \\
 &\;\;\;\;\;=\; 
 \int_0^\infty y^{z-1} \int_{-\infty}^0 L_{s,h}(x,ty) g(t) \rd t \rd y +
 \int_0^\infty y^{z-1} \int_{0}^\infty L_{s,h}(x,ty) g(t) \rd t \rd y
 \nonumber
 \\
 &\;\;\;\;\;=\; \int_0^\infty y^{z-1} \int_{0}^\infty L_{s,h}(x,-ty) g(-t) \rd t \rd y 
 +
 \int_0^\infty y^{z-1} \int_{0}^\infty L_{s,h}(x,ty) g(t) \rd t \rd y
\nonumber 
 \\
 &\;\;\;\;\;=\; 
 \widetilde{L}_{s,h}^-(x,z) \widetilde{g}^-(1-z) + \widetilde{L}_{s,h}^+(x,z) 
 \widetilde{g}^+(1-z),
\nonumber
 \end{align}
 where we denoted $\widetilde{L}_{s,h}^+(x, z)=\cM[L_{s,h}^+(x,\cdot);z]$ and 
 $\widetilde{L}_{s,h}^-(x, z)=\cM[L_{s,h}^-(x,\cdot);z]$.
Similarly, 
\begin{eqnarray*}
 &&\int_{-\infty}^0 (-y)^{z-1} \int_{-\infty}^\infty L_{s,h}(x,ty) g(t) \rd t \rd y 
 \\
 &&\;\;\;\;\;=\; \int_{0}^\infty y^{z-1} \int_{-\infty}^0 L_{s,h}(x,-ty) g(t) \rd t \rd y
 +
 \int_{0}^\infty y^{z-1} \int_{0}^\infty L_{s,h}(x,-ty) g(t) \rd t \rd y 
 \\
 &&\;\;\;\;\;=\; \int_0^\infty y^{z-1} \int_{0}^\infty L_{s,h}(x,ty) g(-t) \rd t \rd y
 +
 \int_0^\infty y^{z-1} \int_{0}^\infty L_{s,h}(x,-ty) g(t) \rd t \rd y
 \\
 &&\;\;\;\;\;=\; \widetilde{L}_{s,h}^+(x,z) \widetilde{g}^-(1-z) + 
 \widetilde{L}_{s,h}^-(x,z) \widetilde{g}^+(1-z).
\end{eqnarray*}
Comparing these expressions with  (\ref{eq:K-Mellin-1})
and (\ref{eq:K-Mellin-2}), we set
\begin{equation}\label{eq:first-equation}
 \widetilde{L}_{s,h}^-(x,z) \widetilde{g}^-(1-z) + \widetilde{L}_{s,h}^+(x,z) \widetilde{g}^+(1-z)
 = \left\{
 \begin{array}{ll}
x^{z-1} \widecheck{K}(zh), & x>0,\\
0, & x<0,
 \end{array}
\right.
\end{equation}
and 
\begin{equation}\label{eq:second-equation}
 \widetilde{L}_{s,h}^+(x,z) \widetilde{g}^-(1-z) + 
 \widetilde{L}_{s,h}^-(x,z) \widetilde{g}^+(1-z) =
 \left\{\begin{array}{ll}
0, & x>0\\
(-x)^{z-1} \widecheck{K}(zh), & x<0.
        \end{array}
\right.
\end{equation}
It is immediate to verify that
solution to equations (\ref{eq:first-equation})--(\ref{eq:second-equation})
is given by
\begin{eqnarray*}
 \widetilde{L}^+_{s,h}(x,z) &=& \frac{\widecheck{K}(zh)}{[\widetilde{g}^+(1-z)]^2 - [\widetilde{g}^-(1-z)]^2}
 \times \left\{\begin{array}{ll}
x^{z-1}\widetilde{g}^+(1-z), & x>0\\
-(-x)^{z-1}\widetilde{g}^-(1-z), & x<0,
               \end{array}
\right.
\\*[4mm]
 \widetilde{L}^-_{s,h}(x,z) &=& \frac{\widecheck{K}(zh)}{[\widetilde{g}^+(1-z)]^2 - [\widetilde{g}^-(1-z)]^2}
 \times \left\{\begin{array}{ll}
-x^{z-1} \widetilde{g}^-(1-z), & x>0\\
(-x)^{z-1}\widetilde{g}^+(1-z), & x<0.
               \end{array}
\right.
\end{eqnarray*}
Applying the inverse Mellin transform 
we obtain 
\begin{align*}
L^+_{s,h} (x, y)&=\frac{1}{2\pi i x} \int_{s-i\infty}^{s+i\infty} \Big(\frac{x}{y}\Big)^z 
\frac{\widecheck{K}(zh) \,\widetilde{g}^+(1-z)}{[\widetilde{g}^+(1-z)]^2-[\widetilde{g}^-(1-z)]^2}\rd z, \;\;\;x>0,\;y>0,
\\*[4mm]
L^+_{s,h} (x, y) &= - \frac{1}{2\pi i x} \int_{s-i\infty}^{s+i\infty} 
\Big(\frac{-x}{y}\Big)^z 
\frac{\widecheck{K}(zh) \,\widetilde{g}^-(1-z)}{[\widetilde{g}^+(1-z)]^2-[\widetilde{g}^-(1-z)]^2}\rd z, 
\;\;\;x<0,\;y>0,
\end{align*}
and 
\begin{align*}
L^-_{s,h} (x, y)&=-\frac{1}{2\pi i x} \int_{s-i\infty}^{s+i\infty} \Big(\frac{x}{y}\Big)^z 
\frac{\widecheck{K}(zh) \,\widetilde{g}^-(1-z)}{[\widetilde{g}^+(1-z)]^2-[\widetilde{g}^-(1-z)]^2}\rd z, \;\;\;x>0,\;y>0,
\\
L^-_{s,h} (x, y) &=  \frac{1}{2\pi i x} \int_{s-i\infty}^{s+i\infty} \Big(\frac{-x}{y}\Big)^z 
\frac{\widecheck{K}(zh) \,\widetilde{g}^+(1-z)}{[\widetilde{g}^+(1-z)]^2-[\widetilde{g}^+(1-z)]^2}\rd z, \;\;\;x<0,\;y>0.
\end{align*}
Comparing these with (\ref{eq:L}) and taking into account that   
$L_{s,h}(x,y)=L_{s,h}^+(x, y)$ when $y\geq 0$ and $L_{s,h}(x,y)=L_{s,h}^-(x, -y)$ when $y<0$ for fixed $x$,
we complete the proof.
\epr

\subsection{Proof of Lemma~\ref{lem:bias}}
Below $c_1, c_2,\ldots$ stand for positive constants depending on $\ell$ only.
By the change of variables, $t=\frac{1}{h}\ln(y/x)$, we have
\begin{eqnarray*}
 && \int \frac{1}{xh} K\bigg(\frac{\ln(y/x)}{h}\bigg) f(y)\rd y -f(x) = 
 \int_{-1}^1 K(t)  [w_x(th) - w_x(0)]\rd t,
 \end{eqnarray*}
where we have denoted $w_x(t):=e^{t} f(xe^{t})$. Since $f\in \sH_{x,r} (A, \beta)$, the function $w_x(\cdot)$
is $\ell$ times continuously differentiable on $[-\ln r, \ln r]$. 
Expanding $w_x(\cdot)$ in Taylor's series around
zero we have for any $t\in [-\ln r, \ln r]$  
\[
 w_x(t)=w_x(0)+\sum_{k=1}^{\ell -1} \frac{1}{k!} w_x^{(k)}(0)t^k + \frac{1}{\ell!} w_x^{(\ell)}(\xi t) t^\ell,
 \;\;\;\xi=\xi(t)\in [0,1].
\]
Therefore if $h<\ln r$ then 
\begin{equation}\label{eq:K-w}
 \bigg|\int_{-1}^1 K(t)[w_x(th) - w_x(0)]\rd t\bigg|\;\leq\;
 \frac{h^\ell}{\ell!} \int_{-1}^1 |t|^\ell |K(t)| \;|w_x^{(\ell)}(\xi th)-w_x^{(\ell)}(0)| \rd t.
\end{equation}
It follows from the Fa\'a~di~Bruno 
formula that for $s>0$
\begin{multline*}
 w_x^{(\ell)}(s) = \sum_{\pi\in \Pi} \Big[(u f(xu))^{(|\pi|)}\Big]_{u=e^s} e^{|\pi|s}
 \\
 = \sum_{\pi\in \Pi} \Big[e^s x^{|\pi|} f^{(|\pi|)}(xe^s) + |\pi| x^{|\pi|-1} f^{(|\pi|-1)}(xe^s)\Big]
 e^{|\pi|s},
\end{multline*}
where the summation runs over the set $\Pi$ of all partitions of the set $\{1,\ldots, \ell\}$,
and $|\pi|$ is the number of subsets in partition $\pi$.
Thus
\begin{multline*}
 w_x^{(\ell)}(\xi th)-w_x^{(\ell)}(0) = 
 \sum_{\pi\in \Pi} e^{(|\pi|+1)\xi th} x^{|\pi|} \big[f^{(|\pi|)}(xe^{\xi th})-f^{(|\pi|)}(x)\big]
 \\
\;\;\;\;+ \sum_{\pi\in\Pi} x^{|\pi|} f^{(|\pi|)}(x)\big[e^{(|\pi|+1)\xi th}-1\big]
+ \sum_{\pi\in\Pi} e^{|\pi|\xi th} |\pi|x^{|\pi|-1} \big[f^{(|\pi|-1)}(xe^{\xi th})-f^{(|\pi|-1)}(x)\big]
\\
+ \sum_{\pi\in\Pi} (e^{|\pi|\xi th}-1)|\pi|x^{|\pi|-1}f^{(|\pi|-1)}(x).
\end{multline*}
In view of $f\in \sH_{x,r}(A,\beta)$  and by elementary inequality $|e^{x}-1|\leq |x|e^{|x|}$, 
\begin{eqnarray*}
 \Big|\sum_{\pi\in \Pi} e^{(|\pi|+1)\xi th} x^{|\pi|} \big[f^{(|\pi|)}(xe^{\xi th})-f^{(|\pi|)}(x)\big]\Big|
 &\leq& c_1 A |x|^\beta |th|^{\beta-\ell} e^{(\beta+1)|th|},
 \\
 \Big|\sum_{\pi\in\Pi} x^{|\pi|} f^{(|\pi|)}(x)\big[e^{(|\pi|+1)\xi th}-1\big]\Big| &\leq&  c_2 A |t h|
 e^{(\ell+1)|th|} \sum_{k=1}^\ell |x|^k,
 \\
 \Big|\sum_{\pi\in\Pi} e^{|\pi|\xi th} |\pi|x^{|\pi|-1} \big[f^{|\pi|-1}(xe^{\xi th})-f^{(|\pi|-1)}(x)\big]
 \Big| &\leq&  c_3 A |th| e^{\ell |th|}\sum_{k=1}^\ell |x|^k,
 \\
 \Big|\sum_{\pi\in\Pi} (e^{|\pi|\xi th}-1)|\pi|x^{|\pi|-1}f^{(|\pi|-1)}(x)\Big| &\leq &
 c_4 A |th| e^{\ell |th|} \sum_{k=0}^{\ell-1} |x|^k.
\end{eqnarray*}
Combining these inequalities and substituting them in (\ref{eq:K-w}) completes the proof.~~\epr

\subsection{Proof of Lemma~\ref{lem:f-y-nearzero}}
We have 
\begin{multline*}
 f_Y(y)= \int_0^\infty \frac{1}{x} f_X(y/x)g(x)\rd x 
 \\
 = \int_0^y \frac{1}{x} f_X(y/x)g(x)\rd x +
 \int_y^\infty \frac{1}{x} f_X(y/x)g(x)\rd x 
 =: I_1 +I_2.
\end{multline*}
Using [G3] for any $y\leq \delta$ and $p\in [0,1)$ we obtain
\begin{multline}
 I_1 \leq C_0 \int_0^y \frac{1}{x} f_X(y/x) x^{-p} |\ln x|^{q} \rd x = C_0 y^{-p} \int_1^\infty 
 t^{p-1} f_X(t) |\ln (y/t)|^{q} \rd t
 \\
 \leq   2^{(q-1)_+} C_0 y^{-p} |\ln y|^{q}\bigg[\int_1^\infty t^{p-1} f_X(t)\rd t + 
 \int_1^\infty t^{p-1}f_X(t) |\ln t|^{q}\rd t\bigg] 
 \\
 \leq 
 c_1 y^{-p} |\ln y|^{q}.
 \label{eq:I-1}
\end{multline}
Since $f_X(t)\leq M$, $\forall t\geq 0$,
\begin{eqnarray}
 I_2 \leq M\int_y^\infty \frac{g(x)}{x}\rd x &\leq& M\delta^{-1} + M\int_y^{\delta} \frac{g(x)}{x}\rd x
 \nonumber
 \\
 &\leq& M\delta^{-1} + C_0M\int_y^{\delta} x^{-p-1} |\ln x|^q \rd x.
 \label{eq:I-2}
\end{eqnarray}
If $p=0$ then  the last integral on the right hand side 
is bounded from above by $|\ln y|^{q+1}$, and 
\[
 I_2 \leq M\delta^{-1} + MC_0 |\ln y|^{q+1}. 
\]
This inequality together with (\ref{eq:I-1}) completes the proof of statement~(i).
\par 
Now we bound the expression on the right hand side of (\ref{eq:I-2}) in the case $p\in (0, 1)$. 
Using the following formula (see, e.g., \cite[616.2]{dwight1961tables})
\[
 \int x^{p-1} (\ln x)^q\rd x = \frac{1}{p} x^p (\ln x)^q - \frac{q}{p} \int x^{p-1} (\ln x)^{q-1}\rd x,\;\;\;
 \forall p\ne 0, \;q\ne -1,
\]
we obtain 
\begin{equation}\label{eq:int-b}
 \int_y^{\delta} x^{-p-1} |\ln x|^q\rd x =\int_{1/\delta}^{1/y} t^{p-1} (\ln t)^q\rd t = 
 \Big[\frac{t^p(\ln t)^q}{p}\Big]_{1/\delta}^{1/y} - \frac{q}{p} \int_{1/\delta}^{1/y} 
 t^{p-1} (\ln t)^{q-1}\rd t.
 \end{equation}
Hence it follows from (\ref{eq:int-b}) that 
\begin{equation}\label{eq:ypp}
 \int_y^{\delta} x^{p-1} |\ln x|^q\rd x \leq \Big[\frac{t^{p} (\ln t)^q}{p}\Big]_{1/\delta}^{1/y}  \leq 
 \frac{y^{-p}}{p}  [\ln (1/y)]^q.
\end{equation}
Thus, we obtain 
\[
 I_2 \leq M\delta^{-1} + C_0Mp^{-1} y^{-p} |\ln y|^q.
\]
Combining this inequality with (\ref{eq:I-1}) we complete the proof of (ii).
\epr

\section*{Acknowledgements}
The authors are grateful to an anonymous referee for careful reading and insightful comments that lead to substantial 
improvements in the paper.

\bibliographystyle{plain}
\bibliography{md-1}

\end{document}